\documentclass[preprint2]{aastex631}
\usepackage{amsmath}
\usepackage{xcolor}

\newcommand{\resubi}[1]{#1}
\newcommand{\resubii}[1]{#1}
\newcommand{\resubiii}[1]{#1}
\newcommand{\resubiv}[1]{#1}

\begin{document}

\title{Quantifying how Surface Complexity Influences Properties of the Solar Corona and Solar Wind}

\author[0000-0002-6478-3281]{Caroline L. Evans}
\affiliation{Department of Astrophysical and Planetary Sciences, University of Colorado Boulder, 2000 Colorado Avenue, Boulder, CO 80305, USA}
\affiliation{Cooperative Institute for Research in Environmental Sciences, 216 UCB Boulder, CO, 80309, USA}
\affiliation{National Solar Observatory, 3665 Discovery Dr, Boulder, CO 80303, USA}

\author[0000-0003-1759-4354]{Cooper Downs}
\affiliation{Predictive Science Inc., 9990 Mesa Rim Road, Suite 170, San Diego, CA 92121, USA} 

\author[0000-0002-9654-0815]{Donald Schmit}
\affiliation{Cooperative Institute for Research in Environmental Sciences, 216 UCB Boulder, CO, 80309, USA}

\begin{abstract}

The Sun's magnetic field is a key driver in coronal heating and consequently solar wind acceleration. Remote measurement of the photosphere provides the magnetic surface boundary condition necessary for data-constrained 3D global coronal models. With one such model, we explore how the spatial resolution of the surface boundary condition influences the global properties of the magnetic field and coronal heating. Using spherical harmonic decomposition, we quantify how three different resolution simulations vary in the low and middle corona. Through examination of the magnetic field, \resubii{the squashing factor}, and the heating rate, we demonstrate that small-scale photospheric magnetic flux enhances heating across spatial regimes. We calculate 40\% more heating in our best resolution simulation as compared to our base resolution. We describe a \resubii{strong} correlation between the structure of the magnetic field and \resubii{structure of the} heating \resubii{rate} in the low corona across resolutions. These results provide key information as to what more efficient, low-resolution models might inherently miss. This can provide context to incorporate the effects of unresolvable features in future modeling efforts.

\end{abstract}

\keywords{Solar corona (1483), Solar coronal heating (1989), Solar photosphere (1518), Solar wind (1534)}

\section{Introduction} \label{sec:intro}

The Sun’s magnetic field affects the heating and structure of the solar corona. At large scales, the photospheric magnetic flux distribution establishes features such as active regions and coronal holes. At smaller scales, the surface flux varies within large-scale topological domains. This introduces \resubii{small-scale variations in magnetic field strength and magnetic field mappings.}

Using the definitions of \citet{Viall2021}, the corona and solar wind have small-, meso-, and large-scale structures. Small-scale magnetic structures and their fluctuations directly influence coronal heating, but unique determination of the mechanism remains elusive. The two most prominent theories are (1) the accumulation and subsequent dissipation of current sheets through Parker braiding or similar nanoflare heating mechanisms \citep{Parker1972, Parker1983, Parker1988, Klimchuk2015} and (2) the propagation and dissipation of magnetohydrodynamic (MHD) waves and their turbulent fluctuations \citep{Velli1993, vanBallegooijen2011}. In either paradigm, footpoint motion generates the Poynting flux necessary for coronal heating. Therefore the magnetic field and its constantly changing, \resubii{small-scale} connectivities drive coronal heating.

At larger scales, the geometry of a closed flux tube (or coronal loop) influences its hydrodynamic properties \citep[see e.g.][]{Mikic2013, Reep2022}. \resubii{For open fields, the relative areal expansion of flux tubes \citep[the expansion factor,][]{WangSheeley1990} describes} the empirical relationship of the magnetic field and flux tube area with the acceleration of plasma as it transitions to the solar wind \citep[see][and references therein]{CranmerWinebarger2019}. The magnetic field defines both open- and closed-field structures. Similarly to heating, this structure affects characteristics such as density and pressure. In turn, density and pressure gradients \resubii{play a large role in determining the characteristics of the solar wind \citep{Parker1958}. But the key physical processes within the corona that} influence the generation, release, and acceleration of the solar wind\resubii{ remain open questions \citep{ViallBorovsky2020}.}

Consequently, accurate modeling of the corona and solar wind requires knowledge of the magnetic field and plasma at a variety of spatial scales. For individual structures, stretched loop models that straighten loops to fit into a Cartesian box are often used. These include the physics that drives small-scale magnetic field fluctuations and their interactions. However, they must simplify loop geometry to capture such high spatial resolution (seen in the work of \citet{Galsgaard1996} and \citet{Breu2022}). As a complement, other high-resolution models like those of \citet{Abbett2007}, \citet{Rempel2017}, and \citet{KanellaGudiksen2019} focus on connecting the convection zone to the low corona. These high-resolution models successfully represent small- and meso-scale physics captured in observations. To expand the modeling domain to the entire corona, a lower spatial resolution must be used within a global solar coronal model. However, the now unresolved features may still be essential for describing coronal heating and solar wind acceleration. For this reason, global MHD models often use `subgrid’ parameterizations as a macroscopic approximation to the underlying, unresolved physical processes \citep[as seen in][]{vanderHolst2014, Usmanov2018, Reville2020}. 

This work utilizes high-resolution global coronal modeling to investigate the multi-scale nature of coronal heating and its links to the solar wind. Ideally, high-resolution global modeling would be the norm, but the computational requirements are generally too large to be practical. We therefore explore how global coronal simulations vary as a function of model resolution. We use a thermodynamic, wave-turbulence-driven (WTD) 3D MHD model. We examine the effect of model resolution by preparing otherwise similar runs at three different effective resolutions. Accurate quantification of the resulting differences reveals the role of small-scale magnetic fields in determining not only the structures and properties of the low corona, but also how these properties transition into the outer corona and heliosphere. We choose to use spherical harmonic decomposition (SHD) so we may quantify heating at different spatial scales and analyze these differences across resolutions. This provides key information about what more cost-efficient, lower-resolution models may not inherently be able to resolve.  

The paper is organized as follows: \S \ref{sec:num} describes our model, methods, and preparation of the data. In \S \ref{sec:results} we offer analysis through SHD of the magnetic field, squashing factor, and heating rate and discuss the results. We make conclusions and offer future steps in \S \ref{ref:conclusions}.


\section{Numerical Methods}\label{sec:num}

\subsection{Magnetohydrodynamic Algorithm outside of a Sphere (MAS)} \label{sec:mas}

To link photospheric structure to the corona and beyond to the solar wind, we use the 3D MHD global coronal model, Magnetohydrodynamic Algorithm outside of a Sphere (MAS). MAS has modeled various regions and phenomena of the Sun \citep{Mikic1999, Lionello2009, Riley2011, Lionello2013, Torok2018}, including eclipses \citep{Mikic2018}. We focus on the data-constrained predictive effort for the 2019 July 2 solar eclipse. This has been previously compared to available observational data \citep{Boe2021, Boe2022, Boe2023}. In this work, we analyze three MAS simulations with identical parameterization but varied effective resolution. 

These simulations use the resistive thermodynamic MHD equations \citep{Lionello2009, Lionello2014, Downs2016, Mikic2018}. In particular, our focus is the inclusion of thermodynamic MHD terms that are incorporated into our energy equation. Within this equation, there are non-ideal terms \citep{Mikic1999, Lionello2002}. These terms encapsulate the dynamics of thermal conduction, radiative losses, and coronal heating. Specifically, MAS uses a Wave-Turbulence-Driven (WTD) approach to model the macroscopic effects of Alfv\'en wave turbulence. This includes separately solving (1) the Wentzel-Kramers-Brillouin (WKB) \resubii{approximation} for the constant propagation of wave energy to accelerate the solar wind and (2) a set of WTD equations to heat the corona through the propagation, reflection, and dissipation of Poynting flux injected at the inner boundary.

As described in \citet{Mikic2018}, MAS implements the WTD heating equations in Elsässer variable representation ($\mathbf{z}_\pm = \delta \mathbf{v} \mp \delta \mathbf{B}/ \sqrt{4\pi\rho}$, where $\rho$ is the density. The flow and magnetic perturbations to the wave are $\delta \mathbf{v}$ and $\delta \mathbf{B}$, respectively). 

\begin{gather}\label{eqn:heat}
    \frac{\partial z_\pm}{\partial t} + (\mathbf{v} \pm \mathbf{v_A}) \mathbf{\cdot \nabla} z_\pm = R_1 z_\pm + R_2 z_\mp - \frac{|z_\mp| z_\pm}{2\lambda_\perp} \\
    R_1 = \frac{1}{4} (\mathbf{v} \mp \mathbf{v_A}) \mathbf{\cdot \nabla} (\mathrm{log}\rho)\\
    R_2 = \frac{1}{2} (\mathbf{v} \mp \mathbf{v_A}) \mathbf{\cdot \nabla} (\mathrm{log}|\mathbf{v_A}|)
\end{gather}

There are three main terms in Equation \resubiv{(\ref{eqn:heat})}: the advective derivative and $R_1$ term describe \resubii{the linear propagation of Alfv\'en waves using the WKB approximation. Rearranging these terms yields the equations of \citet{Jacques1977}.} The $R_2$ term is the WTD self-reflection of Alfv\'en waves.  The remaining term with $\lambda_\perp$ is a phenomenological WTD heating term. Here, $\lambda_\perp$ is the transverse correlation scale and is defined as $\lambda_\perp = \lambda_0 \sqrt{B_0/B}$. In this definition, $\lambda_0$ and $B_0$ are model parameters for the correlation length and reference magnetic field, respectively. The Alfv\'en velocity is $\mathbf{{v_A}}$.

\citet{Downs2016} demonstrate that the wave reflection in MAS depends on gradients of the Alfv\'en speed. They show that the heating of closed coronal structures is highly sensitive to magnetic field variations but relatively insensitive to $\lambda_0$. As the correlation length is more critical to realistic models of the solar wind (see \cite{Lionello2014}), this implementation is compatible with the outer corona and heliosphere. The heating models in MAS and similar models endeavor to describe heating in the solar corona and not the chromosphere or layers below. Here, the chromospheric boundary provides a mass reservoir for the coronal heating problem.
 
This model relies on a full-Sun map of the photospheric magnetic field to drive the simulation. As described in \citet{Boe2021}, the map combines two Solar Dynamics Observatory/Helioseismic and Magnetic Imager (SDO/HMI) synoptic maps \citep{Scherrer2012}. At Carrington Longitude $148\degree$, we  join the definitive synoptic map for Carrington Rotation (CR) 2217 with the Near Real Time synoptic map for CR 2218. This resulting ``parent" map has a resolution of $1800$ $\times$ $3600$ (co-latitude, $\theta$ and longitude, $\phi$).

We create three simulations of this event with the same inner boundary condition of the parent map, but smoothed to each simulation's different \resubi{effective} resolution. All grids are evenly spaced in longitude and \resubi{within} $\pm 65 \degree$ in latitude, but are otherwise non-uniform. The three resolutions (\resubi{which we refer to as} ``medium," ``high," and ``super") of the boundary condition map are $148$ $\times$ $315$, $327$ $\times$ $699$, and $596$ $\times$ $1257$ ($\theta$ $\times$ $\phi$\resubii{, co-latitude $\times$ longitude}). \resubiii{The medium and high simulations have the same mesh ($288$ $\times$ $327$ $\times$ $699$, $r$ $\times$ $\theta$ $\times$ $\phi$); the difference lies in the effective resolution of the boundary condition. In contrast, the super resolution mesh is $445$ $\times$ $569$ $\times$ $1257$.}

To balance computational constraints and minimize differences among our simulations, we strategically order the run sequence and ``remesh" \resubii{following \citet{Linker2024}. This involves remeshing a MAS solution using an integral-preserving interpolation strategy and solving for a new vector potential based on the remeshed non-potential part of the magnetic field.} \resubiii{This preserves magnetic free energy, mass, kinetic energy, and thermal energy, among other quantities. This technique also allows} us to transform a 3D MHD solution from one mesh to another while replacing \resubi{the} boundary condition.
Our high run is constructed from an \resubii{80.3} hour real-time relaxed medium run and relaxed an additional \resubii{44.2} hours. The super run is substantially more computationally expensive (by over a factor of 8)\resubi{. We construct it} from the final state of the high \resubi{run} and relax \resubi{it} \resubii{8.7} additional hours.

\resubiii{Knowing that eight hours of evolution is insufficient time for the solar wind to reach the outer boundary of our simulations and also any new changes in the hydrodynamic state can impart subtle changes on the heliospheric current sheet at a large distance (e.g. $\sim$ 5-30 R$_\odot$),  our analysis purposely focuses only on 3 R$_\odot$ and below. This height is covered by -- at the very least -- several sound crossing times and generally more magnetosonic crossing times. 
We also justified this choice heuristically by looking at the evolution of the total kinetic energy from 1-3 R$_\odot$, which begins to asymptote around 4 hours into the relaxation.}
This is \resubiii{therefore} sufficient to achieve a quasi-relaxed state by at least 3 R$_\odot$. 

\resubiii{To summarize:} as a result of the computational cost for the super simulation, \resubi{the super} steady-state solution is only reliably converged in the low and middle corona \resubi{\citep[see the definition of][of approximately 1.5 - 6 R$_\odot$]{West2023}.}

We therefore analyze the low and middle corona \resubiii{(at 3 R$_\odot$)} in all three simulations. Examination of these regimes is sufficient to understand the \resubi{resulting} density and pressure profiles of the solar wind \citep{DeForest2018, Seaton2021, Chitta2023}.

\subsection{Spherical Harmonic Decomposition}\label{sec:SH}

\resubi{Historically, magnetic field reconstructions based on observations and within models depend on spherical harmonics (SHs), making them inherent to our understanding of the Sun \citep{Schatten1969}. For example, potential field solutions combine data and SH expansion algorithms \citep{Schrijver2003}\resubi{. Additionally,} SHs are the mathematical basis for analysis of magnetic fields and related vector and scalar quantities \citep{DeRosa2012, Hathaway2000, Caplan2021, Yoshida2023}. }

We \resubi{therefore choose to use} spherical harmonic decomposition (SHD)\resubi{ for our quantification efforts. We utilize} routines available in pyshtools.\footnote{https://shtools.github.io/SHTOOLS/index.html} This package is a Python wrapper to the FORTRAN 95 library SHTOOLS \citep{Wieczorek2018} and has been updated to include compatibility with the C++ library DUCC (Distinctly Useful Code Collection). 

We $4\pi$ normalize and exclude the Condon-Shortley phase factor. As a result, our data is purely real-valued and we use the real SHs. 

\begin{gather} \label{eqnSH}
 f(\theta,\phi) = \sum_{l=0}^\infty \sum_{m=-1}^l f_{lm} Y_{lm}(\theta,\phi) \\
 Y_{lm} = 
\begin{cases}
P_{lm}(\text{cos}\theta)\text{cos}(m\phi) & \text{if } m \geq 0,\\
P_{lm}(\text{cos}\theta)\text{sin}(|m|\phi)  & \text{if } m < 0.
\end{cases}\\
\frac{1}{4\pi} \int_\Omega f^2(\theta,\phi) d\Omega = \sum^\infty_{l=0} S_{ff}(l)\\
S_{ff}(l) = \sum_{m=-l}^l f^2_{lm} \label{eqnTPS}
\end{gather}

Here, $f(\theta, \phi)$ represents data with angular dependence, $Y_{lm}$ are the real SHs, $f_{lm}$ are the coefficients of the SHs, $P_{lm}$ represents the normalized Legendre polynomials, and $S_{ff}(l)$ is the total power \resubi{spectrum}. The total power \resubi{spectrum} removes $m$- or order-dependence and depends only on $\ell$, the SH degree.

We apply SHD to the steady-state, relaxed 3D MHD simulations described in \S \ref{sec:mas}. Without any time evolution in our boundaries, the symmetry and placement of structures are not immediately relevant. We \resubi{therefore choose to calculate and analyze the total power spectrum. This gives the} power of structure \resubi{(per $\ell$) } to quantify the import of \resubi{relevant} spatial scales.

\resubi{Additionally}, Parseval's theorem is valid to examine the total power spectrum:
\begin{equation} \label{parseval}
 \frac{1}{4\pi} \int_\Omega f^2(\theta, \phi) d\Omega = \sum_{lm} C^2_{lm} \frac{N_{lm}}{4\pi} 
\end{equation}
where $N_{lm}$ is the normalization and $C_{\ell m}$ are the coefficients of the total power spectrum for each SH. With $4\pi$-normalization, Parseval's theorem indicates that the normalized angular integral of the square of the data is equal to the sum of the total power coefficients, squared. 

To perform SHD, we must interpolate our non-uniform grid to a uniform grid. We use Piecewise Cubic Hermite Interpolating Polynomials (PCHIP) \citep{Fritsch1984}. This algorithm prevents ringing and accounts for extrema through preservation of monotonicity, as opposed to pure cubic spline interpolation. We interpolate each radial slice to a resolution of 1256 $\times$ 2512. \resubi{This results in $\ell_{\mathrm{max}} = 627$.} 

We choose our $\ell_{\mathrm{max}}$ to ensure no computational errors from pyshtools itself, as its SHD speed and accuracy is proven through $\ell=2800$ \citep{Wieczorek2018}. Further, $\ell_{\mathrm{max}} = 627$ encapsulates a range of structures, including those on the order of giant cells (peaking around $\ell \sim 10$), supergranules ($\ell \sim 120$), and mesogranules ($\ell \sim 600$) \citep{Hathaway2000, Hathaway2015, Hathaway2021}. 

There is an inverse relationship between $\ell$-space and spatial scales\resubi{. The} constant of proportionality varies geometrically depending on the radius of the surface in question as seen in \citet{Gelabert1989} and \citet{Luo2023}. \resubi{Regardless, lower} $\ell$ values correspond to the larger-scale structures and higher $\ell$ values correspond to smaller ones. \resubii{Whereas the spatial scale in e.g. kilometers will evolve with changing radius, the angular size corresponding to $\ell_{\mathrm{max}} = 627$ is $\sim 0.6 \degree$ (via $\frac{2 \pi}{\ell}$).}

We apply SHD on our prepared data and the absolute value of the square root of our prepared data. We maintain real SHs with the inclusion of the absolute value. \resubi{From Equation \resubiv{(\ref{parseval})}, the units of the power spectra are in units of the quantity, squared. As an example, this means the units of the total power spectrum of the square-rooted heating rate has units of $\mathrm{erg/(cm}^{3} \mathrm{s})$, while the units for the total power spectrum of the heating rate is} $\mathrm{erg^2/(cm}^{6} \mathrm{s^2})$. 
This allows us to quantify the impact of specific spatial regimes in physical units, as seen in \S \ref{sec:results}. It is worth noting that taking the absolute value of a signed quantity can result in a bias in the total power spectrum toward smaller spatial scales. This spreads the pure spectral signature and can disperse power over nearby $\ell$ values. \resubii{By applying the exact same steps for SHD to each simulation and using the same logarithmic-spaced bins to group large ranges of $\ell$ throughout \S \ref{sec:results}, we }minimize this effect.  


\section{Results}\label{sec:results}
\subsection{Radial Magnetic Field in the Photosphere and Middle Corona}\label{sec:mag}
\begin{figure*}[hbt!] 
   \centering

   \includegraphics[scale=0.58]{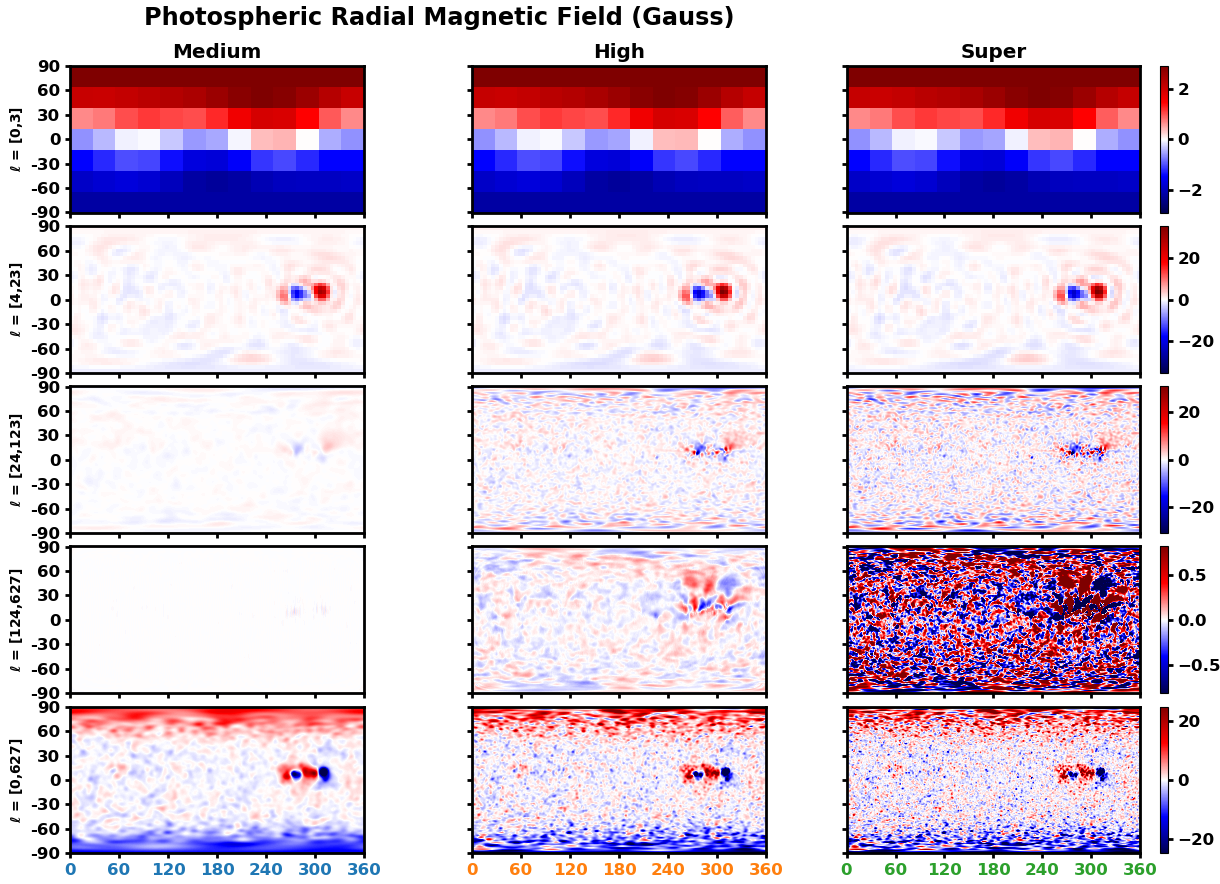}
   
    \caption{After applying SHD on the photospheric boundary condition synoptic map (left: medium, middle: high, right: super), the SHD is re-composed into data for logarithmic-spaced bins in $\ell$. Units are in Gauss and \resubi{relative for each} row. First row: Regridded data for $\ell = [0, 3]$, with identical features across resolution. Second row: Regridded data for $\ell = [4, 23]$, largely corresponding to the active region. Third row: Regridded data for $\ell = [24, 123]$, with diminishing magnetic field structures in the medium simulation. Fourth row: Regridded data for $\ell = [124, 627]$, where the magnetic structures in the super simulation are more prominent than both the high and medium resolutions. Fifth row: the SHD back to gridded data for the total range of $\ell$ values, which are visual matches to the three upper left panels of Figure \ref{fig:br_combo_1}.}
    \label{fig:br_SHViz_1}
\end{figure*}

We first apply SHD to the \resubi{inner magnetic} boundary condition of each simulation. As detailed in \S \ref{sec:mas}, the boundary conditions are \resubi{derived from the same parent map. The only difference is that we perform flux-preserving smoothing to create the effective }medium, high, and super resolutions. \resubi{These maps} constrain the radial magnetic field at the photosphere. For \resubi{reference, in} a current-free potential electromagnetic field, each $\ell$-value corresponds to multipole expansion terms (see \cite{Altschuler1969} and references therein). 

In Figure \ref{fig:br_SHViz_1}, after performing SHD, we \resubi{recompose the} data using only the $\ell$ values for four different logarithmic-spaced bins: [0, 3], [4, 23], [24, 123], and [124, 627]. The first bin contains the $\ell$ values of the monopole, dipole, quadrupole, and octupole moments of the magnetic field. In row 1, we show that the largest structures are identical across resolutions (for medium (left), high (middle), and super (right) simulations). Row 2 highlights the active region around $0 \degree$ latitude and $330 \degree$ longitude and still maintains similar structures across resolutions. The differences among boundary conditions \resubi{become} discernible in the third and fourth bins (row 3, $\ell = [24, 123]$ and row 4, $\ell = [124, 627]$, respectively). The ``salt and pepper" background structures throughout the photosphere are evident\resubi{. Furthermore,} the magnitude of the magnetic field in the super and high simulations are greater than that of the medium resolution. The last row of this figure represents the radial magnetic field fully transformed back into data using \resubi{all $\ell$ ([0, 627])}. 

Based on our construction of the changing boundary condition, these results are expected. We can now interpret the associated total power spectra (Equation \resubiv{(\ref{eqnTPS})}).

As detailed in \S \ref{sec:SH}, the \resubi{relationship between the spatial scale and $\ell$-value} is inversely proportional\resubi{. Therefore, analysis of the total power spectrum, which is dependent on $\ell$,} focuses on the quantity of structures at a given spatial scale. We use ``low” and ``high” to refer to the numeric values of $\ell$ and ``large” and ``small” to refer to the respective spatial scales. 

\begin{figure*}[hbt!] 
   \centering
    \includegraphics[scale=1]{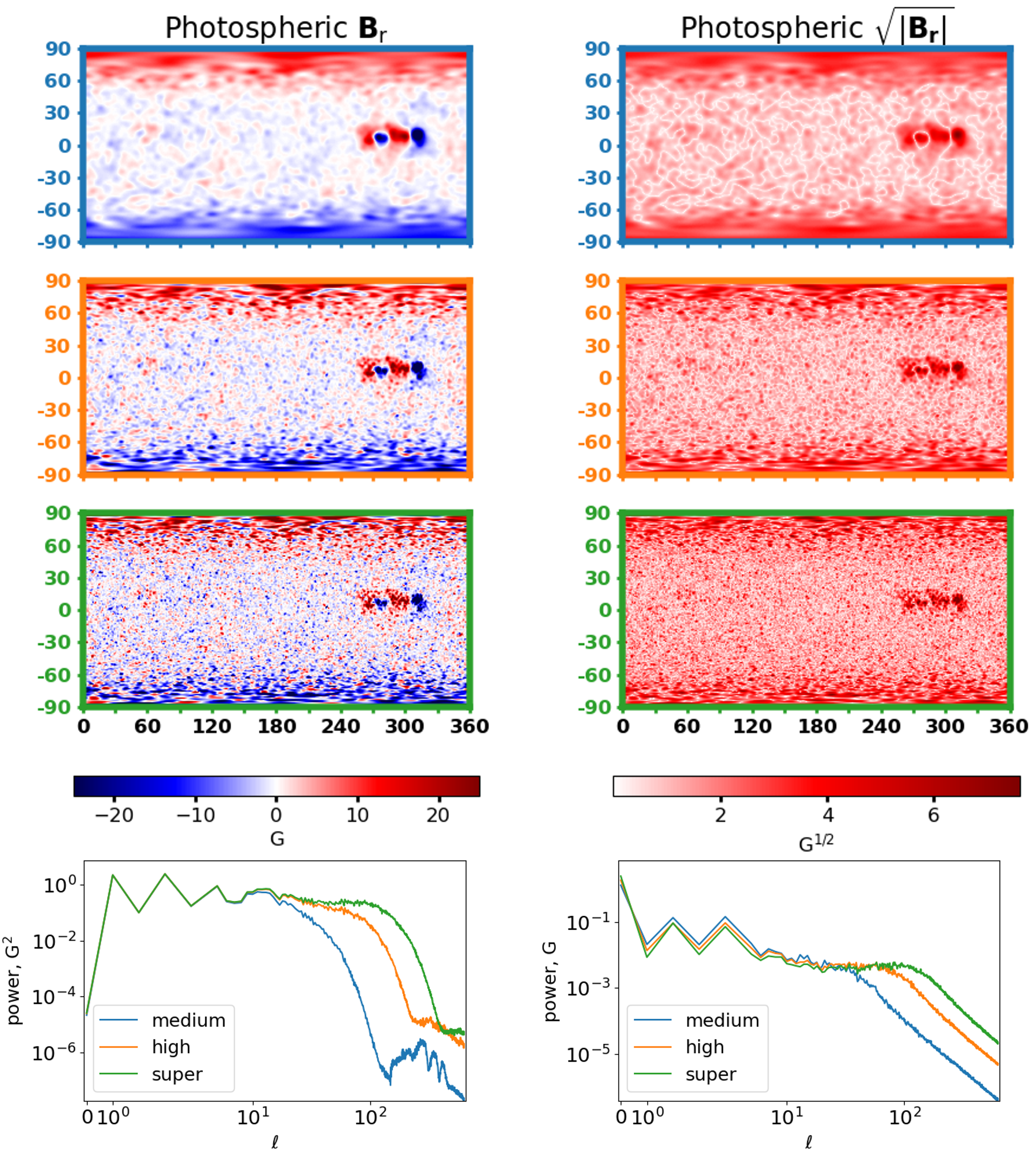}

    \caption{Three upper left panels: the photospheric radial magnetic field (Gauss) of each simulation (in Carrington longitude and latitude): medium (top, blue), high (middle, orange), and super (bottom, green). Upper right panels: \resubi{similarly,} the square root of the unsigned flux ($\sqrt{\mathrm{G}}$) of the photospheric radial magnetic field. Bottom panels: the associated total power spectra in units of the inputted data, squared. The power of the super simulation remains greater than the high and medium simulations at high $\ell$ values\resubi{. This indicates more power in the super simulation} at these smaller spatial scales\resubi{. This is a quantification of the qualitative differences show\resubiv{n} in }the upper panels.}
    \label{fig:br_combo_1}
\end{figure*}

Figure \ref{fig:br_combo_1} \resubi{presents our analysis for the photospheric (1 R$_\odot$) boundary conditions for each simulation. In the left column, we include }the visualization (upper panels) and total power spectra for each resolution (bottom row) of the radial magnetic field\resubi{. We show the same analysis in the right column for the} square root of the absolute value of the radial magnetic field. We include the \resubi{right column} for three reasons. (1) We can compare the unsigned radial magnetic flux in relationship to the radial magnetic field. (2) The sum of these power spectra is equivalent to the integral of the unsigned flux (see Equation \resubiv{(\ref{parseval})}), connecting a physical quantity to the total power spectra. (3) We replicate this unit analysis for the heating rate, another positive definite scalar, in \S \ref{sec:heat}.

The total power spectra of the radial magnetic field (bottom left) quantifies \resubi{what we see qualitatively in the upper panels of Figure \ref{fig:br_combo_1}}. For low $\ell$, the largest-scale structures are identical \resubi{across} simulations. With increasing $\ell$, the power of the super resolution remains larger than that of the high and medium resolutions. \resubi{This indicates that the} super simulation retains more small-scale structures than the high and medium. \resubi{This reflects the smoothing of the parent map.}

\resubi{We also note that} the monopole moment is significantly smaller than the \resubi{power for} subsequent $\ell$-values in the left total power spectra in Figure \ref{fig:br_combo_1}\resubi{. Meanwhile, the power for $\ell=0$ in the total power spectra on the} right is much greater. This difference is expected; the monopole term of the unsigned flux is not required to be zero.

\begin{figure*}[hbt!] 
   \centering
   \includegraphics[scale=1]{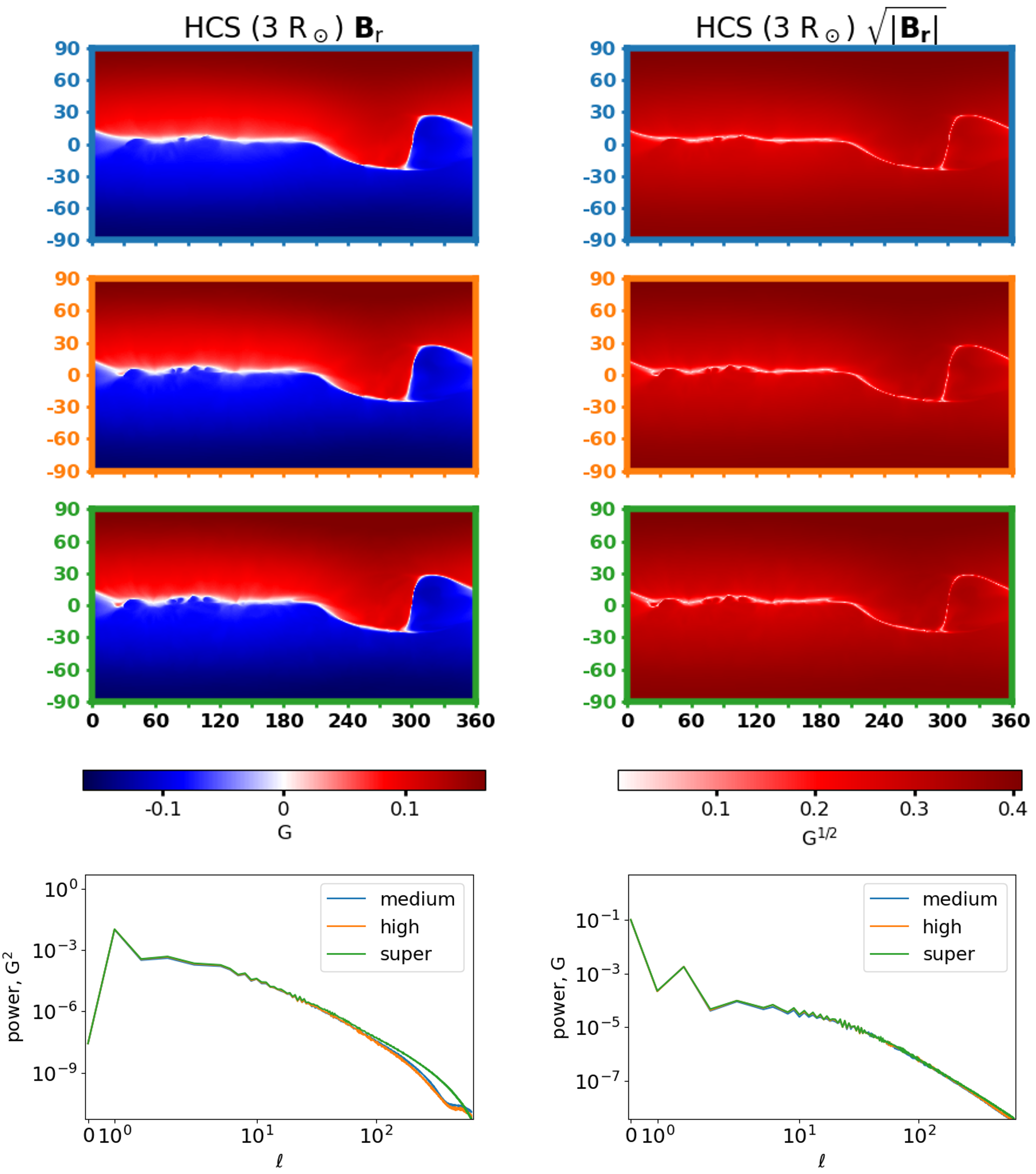}
    \caption{As in Figure \ref{fig:br_combo_1}, the SHD and associated power spectra of the radial magnetic field (G) and square root of the unsigned flux (G$^{1/2}$), but now at 3 R$_\odot$. The overarching large-scale \resubii{location of the HCS (i.e. the position of the inversion line where B$_r$ = 0) suggests that} there should be similar structure across resolutions. Their similar power spectra confirms this. }
    \label{fig:br_combo_3}
\end{figure*}

For the square root of the unsigned flux, the power at the lowest $\ell$ values of the medium and high resolutions have a greater magnitude than that of the super resolution. \resubi{As discussed in \S \ref{sec:SH}, this is the result of} taking the absolute value of a signed quantity. The relative structure, nevertheless, matches. Similarly to the power spectra of the magnetic field, the power of the super simulation remains greater in magnitude for more $\ell$ values than for high or medium. \resubi{The} largest structures influence the power spectra in the lower resolution more than in the higher resolution.

We repeat this analysis \resubi{in Figure \ref{fig:br_combo_3}} for another region of interest in the corona, the heliospheric current sheet (HCS). \resubi{We select a height of 3 R$_\odot$ to analyze. Here}, the three simulations look visually similar for both the magnetic field and square root of the unsigned flux (upper panels) and are nearly identical in the total power spectra (lower panels). 

\begin{figure*}[hbt!] 
     \centering
    \includegraphics[scale=.41]{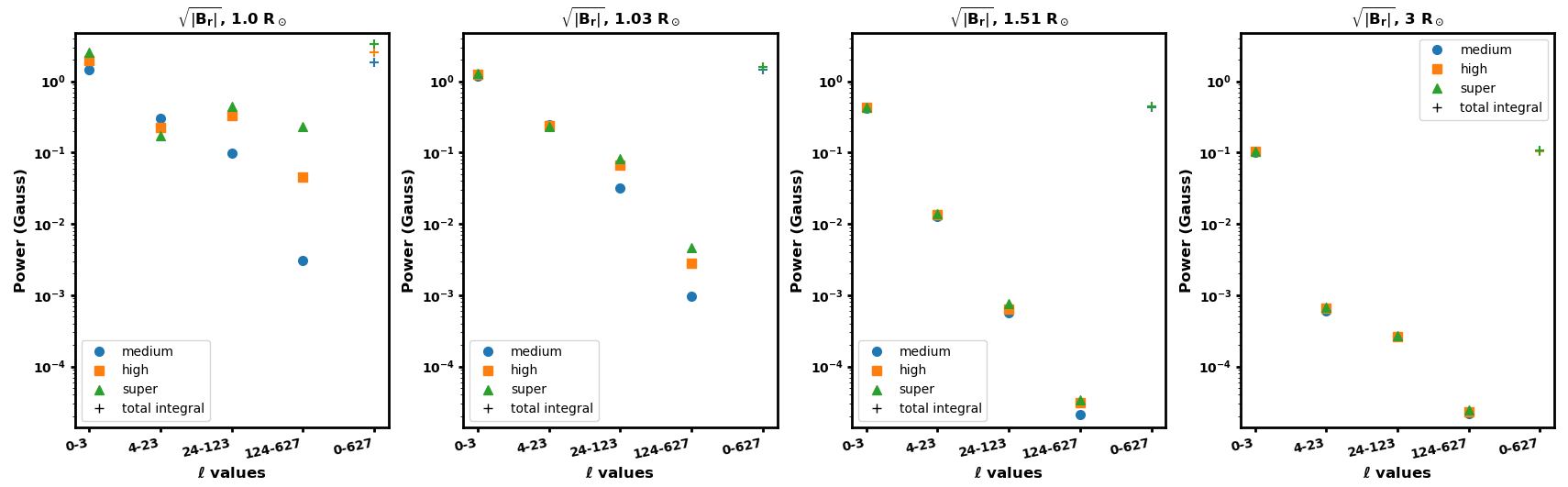}
    \caption{The power spectra of the square root of the unsigned radial magnetic field summed by logarithmic bins for each slice (1.0 R$_\odot$, 1.03 R$_\odot$, 1.51 R$_\odot$, 3 R$_\odot$) examined. There are noticeable magnitude difference in the lowest $\ell$ values in the photosphere (left, 1.0 R$_\odot$)\resubi{. In} the low corona (center left, 1.03 R$_\odot$), the differences in the largest spatial scales have diminished while \resubi{they have persisted} in the smallest spatial scales. As the simulation approaches the heliospheric current sheet (center right, 1.51 R$_\odot$ and right, 3 R$_\odot$), the power of the structures at each spatial regime have become nearly identical.}
    \label{fig:br_slices}   
\end{figure*}

To characterize this evolution with changing height, we use \resubi{the same logarithmic-spaced bins }([0, 3], [4, 23], [24, 123], and [124, 627]) to examine the power at each spatial scale\resubi{. For the given bins, we sum the power spectra} of the square root of the unsigned flux and present these results in Figure \ref{fig:br_slices}. \resubi{Consequently, the} units for the power in each bin is in Gauss (see Equation \resubiv{(\ref{parseval})}). In the photosphere, the features of the power spectra of Figure \ref{fig:br_combo_1} are reflected in the leftmost panel of Figure \ref{fig:br_slices}\resubi{. The} large difference in the power of the smallest scale structures are present, as is the mild excess power in the medium simulation in the $\ell$ = [4, 23] bin. However, the overall power of the super over the high and medium simulations is evident in the total power (plotted \resubi{rightmost as} a + marker), where the super simulation has more flux as compared to both the high and medium resolutions. 

\resubi{With increasing height, the} differences in the power of the lowest $\ell$ values diminish. In the low corona (1.03 R$_\odot$), the power of the highest $\ell$ bins varies \resubi{across} simulations, but \resubi{values of the }initial bin ([0, 3]) remain greatest. The magnitude of \resubi{this} bin nearly matches that of the [0, 627] bin. This trend continues into the middle corona (1.51 and 3 R$_\odot$), but the power spectra and therefore the simulations homogenize to the nearly uniform large-scale structure of Figure \ref{fig:br_combo_3}. Additionally, the magnitude of the power spectra decreases as the \resubi{strength of the }radial magnetic field decreases. 
 
\resubi{This progression toward nearly identical total power spectra suggests that} SHD \resubi{effectively} captures the structural properties of the radial magnetic field in our simulations. \resubii{This is consistent with the fact that small-scale photospheric flux should not impact the open flux \citep{WangSheeley2002, Caplan2021}. This makes intuitive sense given that high order terms of a multipole expansion decay much more quickly than the low order terms, which limits the radial influence of small-scale magnetic features on the Sun. } \resubi{However, there exist} other quantities that \resubi{preserve} the differing structure of the photosphere into the middle corona. 

\resubii{Two such quantities are the squashing factor and the heating rate. The former characterizes magnetic flux domains and therefore the resulting (in this case) radial connectivities of the magnetic field. Including more small-scale magnetic flux will change the flux domains and alter the surface to middle coronal connectivities, which we expound on in \S \ref{sec:Q}. The heating rate depends on multiple properties of the coronal plasma (Equation \resubiv{(\ref{eqn:heat})} and we analyze the heating rate differences in \S \ref{sec:heat} and \S \ref{sec:vol}.}

\subsection{Squashing Factor in the Photosphere and Middle Corona} \label{sec:Q}

\begin{figure*}[hbt!] 
    \centering
    \includegraphics[scale=1]{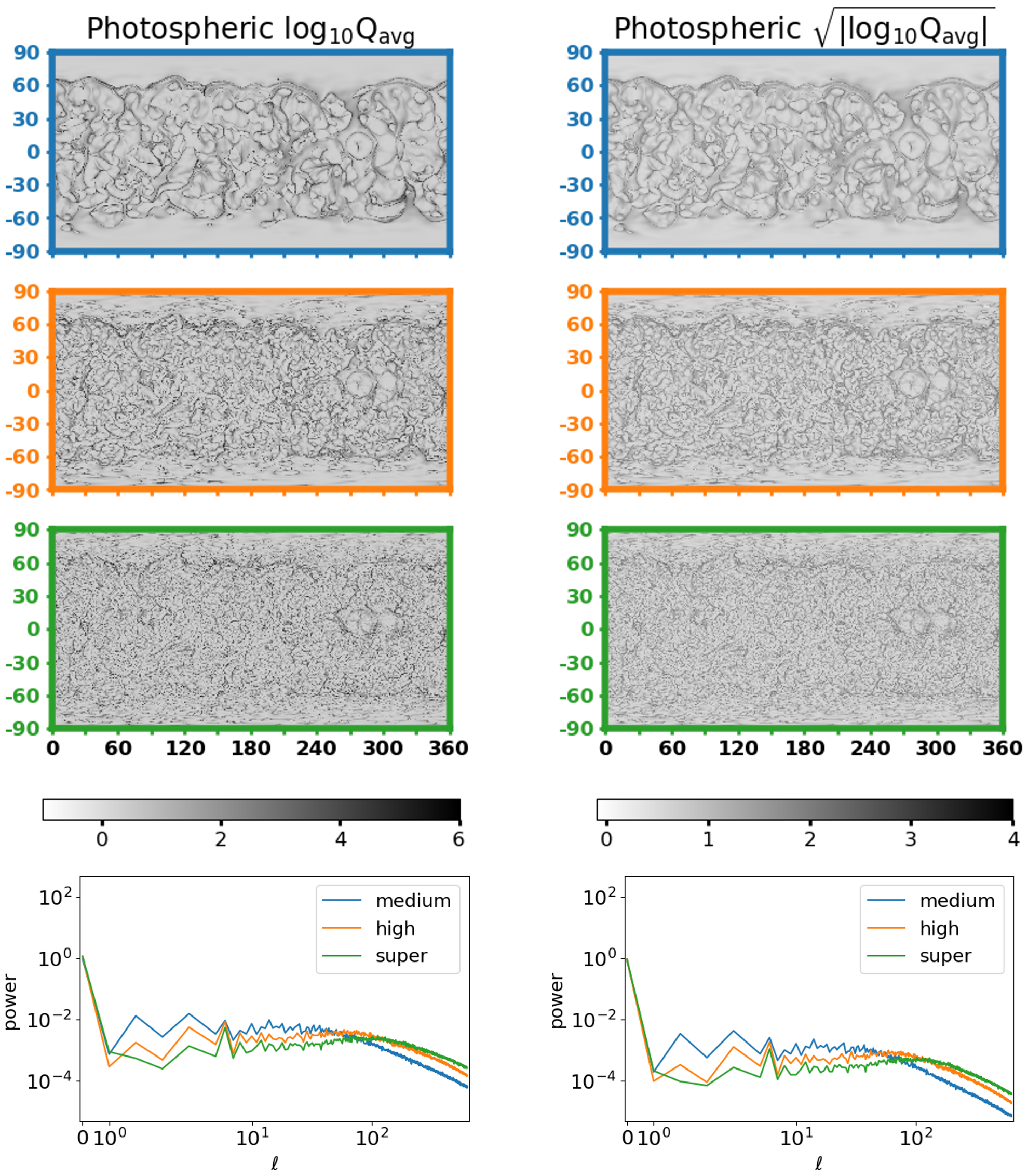}
    \caption{Constructed analogously to Figure \ref{fig:br_combo_1}, the visualization and total power spectra for all three resolutions of the photospheric log$_{10}$Q$_{\mathrm{avg}}$\resubi{ (left)} and $\sqrt{|\mathrm{log}_{10}\mathrm{Q}_{\mathrm{avg}}|}$ \resubi{(right)}. For this scalar quantity, the structures remain nearly identical in the upper left and upper right panels\resubi{. This is further shown} in the similar shapes of the power spectra in the lower panels. The medium power spectrum having more power at low $\ell$ values is a reflection of the fractal nature of the squashing factor\resubi{. A possible example of this effect is} the band structures around $\pm 60 \degree$ latitude.}
    \label{fig:Q_combo_1}
\end{figure*}

\begin{figure*}[hbt!]
    \centering
    \includegraphics[scale=1]{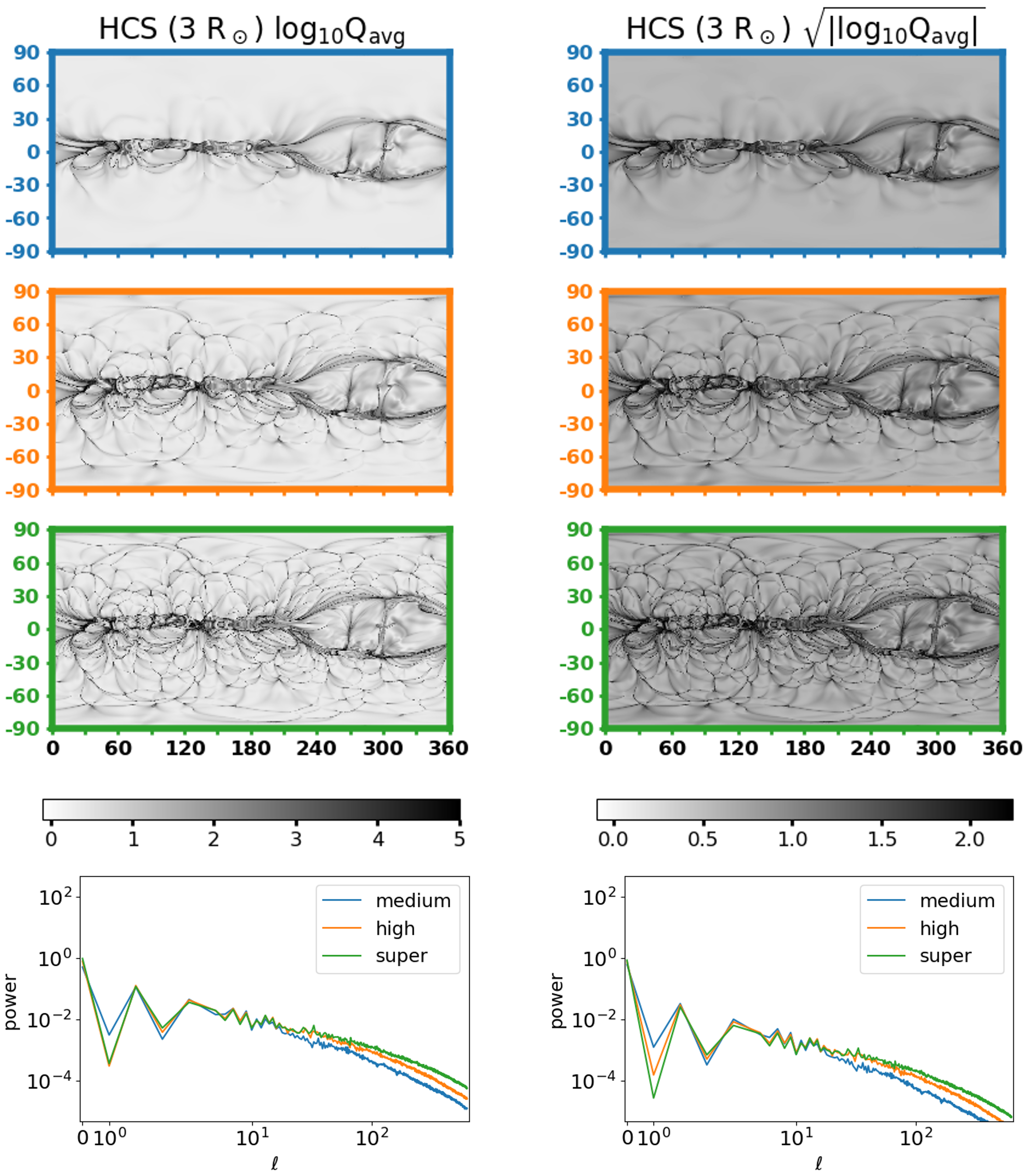}
    \caption{The visualization and SHD power spectra of the squashing factor\resubi{ (left)} and the square root of its absolute value\resubi{ (right) at the} HCS height of 3 R$_\odot$. Unlike Figure \ref{fig:br_combo_3}, there are still differences in the structure of the upper panels that are reflected in the power spectra of the lower panels. There are a greater number of domain cells extending from the HCS. This greater number of small-scale structure appears in the separation of the power spectra at high $\ell$ values in the lower panels.}
    \label{fig:Q_combo_3}
\end{figure*}

The squashing factor Q is a topological measure \resubi{describing the} distortion of an infinitesimally small flux tube mapped from a given surface \citep{Titov2002, Titov2007}. 
Q is a connectivity mapping \resubi{where high Q-lines} subdivide magnetic flux domains\resubi{. The} structure of the surface flux distribution can \resubi{therefore} be traced to any height. With Q, potential locations for magnetic energy storage and release (e.g., flux ropes and reconnection) can be identified \citep{Sweet1969, Lau1990}.

High values of Q indicate the likely presence of quasi-separatrix layers (QSLs) \citep{Priest1995}. QSLs are probable sites for current sheet formation and therefore also the accumulation of free magnetic energy \citep{Priest1996, Longcope2001}. Moreover, QSLs tend to be the geometric representation of the phase space condition of separatrices \citep{Titov2007}, in the sense that QSLs often demarcate different physical outcomes for adjacent parcels of plasma.

\begin{figure*}[hbt!]
     \centering
    \includegraphics[scale=0.41]{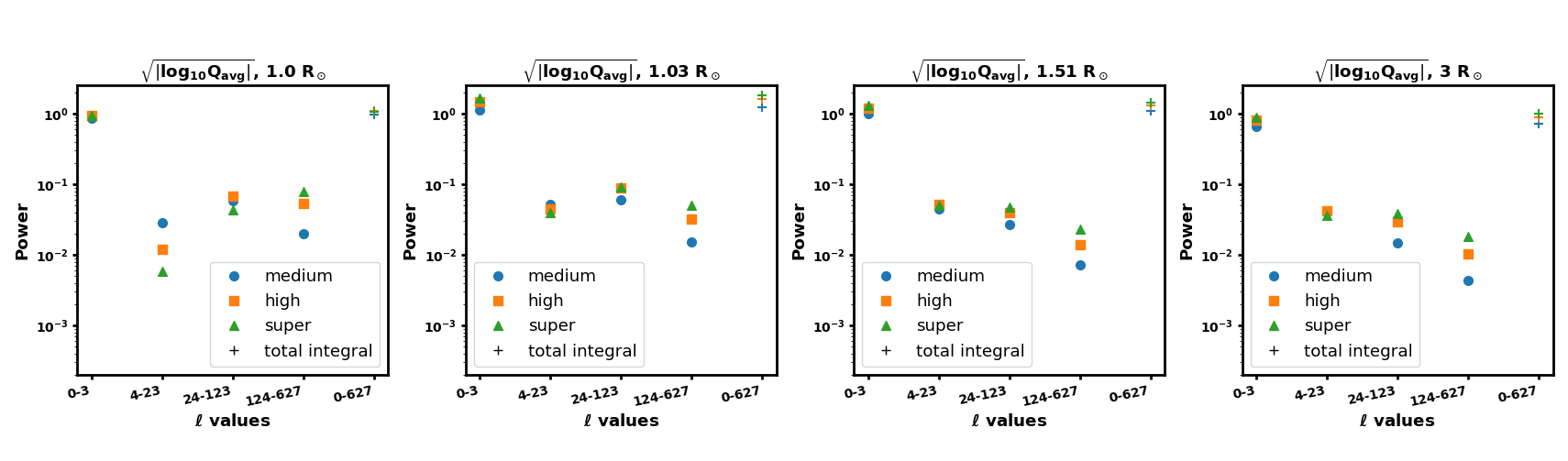}

    \caption{The power spectra \resubi{for} the square root of the absolute value of log$_{10}$ Q for each slice (1.0 R$_\odot$, 1.03 R$_\odot$, 1.51 R$_\odot$, and 3 R$_\odot$) of interest\resubi{. This is constructed in the same manner as} Figure \ref{fig:br_slices}. The distinct difference in flux domains across simulations is maintained across each slice, evidence of the influence of small-scale photospheric magnetic flux.}
    \label{fig:Q_slices}   
\end{figure*}

\citet{Antiochos2011} and \citet{Titov2011} examine large-scale QSLs for smooth flux distributions to \resubi{relate photospheric flux segmentation to the HCS.} Our work, particularly for the high and super simulations with ``salt and pepper" features in the photosphere, will inherently have many more small-scale QSLs\resubi{. These, in turn,} will have signatures in the HCS (see also \citet{Caplan2021}).

At QSLs, both field line mappings and flux-tube properties greatly diverge. Quantification of the structure and scales of these flux domains is a measurement of the complexity of the magnetic field. \resubi{This} reveal how the \resubi{topological} information of the photospheric magnetic field propagates into the corona.

\resubi{With the same implementation as Figure \ref{fig:br_combo_1}, we} apply SHD to the squashing factor in the photosphere. We present these results in Figure \ref{fig:Q_combo_1}. 
As this is a scalar quantity, the similar profiles between $\mathrm{log}_{10}\mathrm{Q}_{\mathrm{avg}}$ and $\sqrt{|\mathrm{log}_{10}\mathrm{Q}_{\mathrm{avg}}|}$ are expected; taking the absolute value and square root will not substantially change the structure of the magnetic flux domains. \resubii{As mentioned in \S \ref{sec:mag}, taking the logarithm of a quantity can spread its spectral signature. However, our identical treatment of each simulation and using logarithmic-spaced bins in $\ell$ (discussed later in Figure \ref{fig:Q_slices}) minimizes this effect.}
In contrast to the results for the radial magnetic field, the power at low $\ell$ of the medium resolution power spectrum is consistently greater than that of the high and super resolution. \resubi{As} the total power spectrum depends only on the degree $\ell$ of the spherical harmonic and not the order $m$, a representative feature that might explain the extra power in the medium resolution is the bands at $\pm 60\degree$ Carrington latitude in the upper panels of Figure \ref{fig:Q_combo_1}.
These bands in the medium data are increasingly obscured in the high and super \resubi{runs}. While the bands exist in the high and super simulations, \resubi{a} fractal nature \resubi{appears because} they are composed of smaller flux domains, which is notable in the greater magnitude of power at \resubi{higher} $\ell$ values. 

While the power spectra appear different in Figure \ref{fig:Q_combo_1}, their total power is nearly identical (see the + marker in the leftmost panel of Figure \ref{fig:Q_slices} and subsequent discussion). \resubi{There is a noticeable} power increase \resubi{for} lower $\ell$ values \resubi{in} the medium simulation as compared to the high and super simulations\resubi{. This} is a reflection of the diminished power \resubi{of} the medium resolution of the radial magnetic field in Figure \ref{fig:br_combo_1}. Q, as a measure of the complexity of the magnetic field, can resolve spatial scales smaller than the magnetic field. But, it cannot create structure where there is none.

The total power spectra represent the increasingly fractured nature of the magnetic flux domains traced out by the squashing factor. The spatial scales of these boundaries decrease as the scales of the magnetic flux defining them also decreases\resubi{. This raises} the amount of power at the smallest spatial scales. This increase \resubi{in power at} higher $\ell$ highlights the increasing complexity in connectivities that result from increasing the magnetic field resolution. In turn, the medium simulation will inherently have more power at larger spatial scales.

In Figure \ref{fig:Q_combo_3}, we again examine the simplified structure of the HCS (centered around $0 \degree$ latitude) at 3 R$_\odot$. All three simulations show matching structures, which the similar total power spectrum for each captures. \resubi{Here, the} super simulation \resubi{once again has} more power than the high and medium for higher $\ell$ values. This is a quantitative representation of the greater number of magnetic flux domains that extend from the surface to the HCS with increasing resolution. With more fragmented flux in the photosphere, there are more possible magnetic flux domains. More magnetic flux domains at the height of the HCS indicates that the squashing factor captures the complexity of the photosphere throughout the simulation. 

We compare the evolution of these magnetic flux domains with application of Equation \resubiv{(\ref{parseval})} in Figure \ref{fig:Q_slices}. Similarly to Figure \ref{fig:br_slices}, \resubi{we sum to find the power within} four logarithmic-spaced bins for radial heights 1, 1.03, 1.51, and 3 R$_\odot$. We also present the overall power as the fifth data point (with a + marker). Across radial heights, the total power in $\sqrt{\mathrm{|log_{10}Q}|}$ of the super resolution exceeds (or matches) that of the high and medium runs and nearly matches the power of the largest spatial scales (the first logarithmic bin, $\ell = [0,3]$), similarly to Figure \ref{fig:br_slices}. 

\resubi{The greater amount of power in the medium resolution at larger scales -- as seen in Figure \ref{fig:Q_combo_1} -- appears in the leftmost panel of Figure \ref{fig:Q_slices}.} More notably, \resubi{for all slices, the super simulation has more power }in the fourth logarithmic bin ($\ell = [124,627]$) than high and medium. This is a strong contrast to the equivalent analysis of the radial magnetic field in Figure \ref{fig:br_slices}. Whereas the power in even the highest $\ell$ value bins approach nearly identical values in the middle corona, the power of the highest $\ell$ value bins for the squashing factor do not.

\begin{figure*}[hbt!] 
    \centering
    \includegraphics[scale=1]{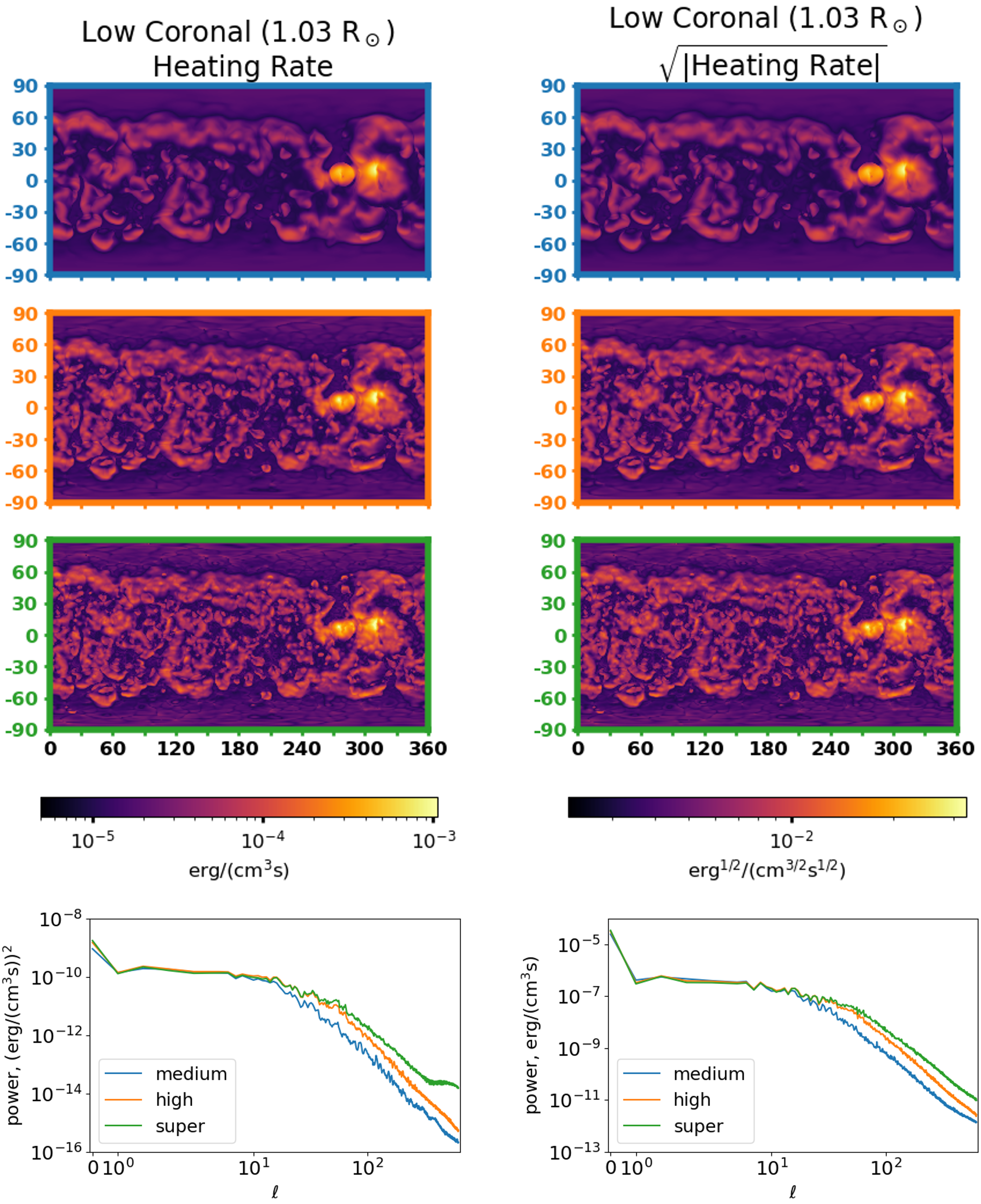}

    \caption{\resubi{We perform} SHD on the heating rate and the square root of its absolute value for our fiducial low corona height of 1.03 R$_\odot$. The subtle increasing background values (\resubiv{in the data visualization of the upper panels, in} purple) indicate heating from the \resubi{improved resolution of the} photosphere has raised the overall heating. The \resubi{super resolution power spectrum confirms a} slightly larger $\ell=0$ term. \resubi{See Figure \ref{fig:heat_slices} for further quantification.}}
    \label{fig:heat_combo_1.03}
\end{figure*}

\resubi{This matches the qualitative expectation that} within the middle corona, the radial magnetic field coalesces (see Figure \ref{fig:br_combo_3})\resubi{ and }the squashing factor does not \resubi{(Figure \ref{fig:Q_combo_3})}. Consequently, the underlying topology of the magnetic field provides an alternative method to quantify the effects of increasing resolution from the photosphere into the middle corona. By increasing the resolution of the photosphere, \resubi{the dipole field of the HCS increasingly fragments into distinct flux domains.} This is evidence that the small-scale photospheric flux elements influence the connectivities of the solar wind.

\subsection{Heating in the Low and Middle Corona} \label{sec:heat}

We \resubi{apply SHD to} each simulation's coronal heating. We pick two fiducial cases for each region\resubi{ of interest}: 1.03 R$_\odot$ for the low corona and 3 R$_\odot$ as the same middle coronal height analyzed \resubi{previously}. 

\begin{figure*}[hbt!] 
   \centering
    \includegraphics[scale=1]{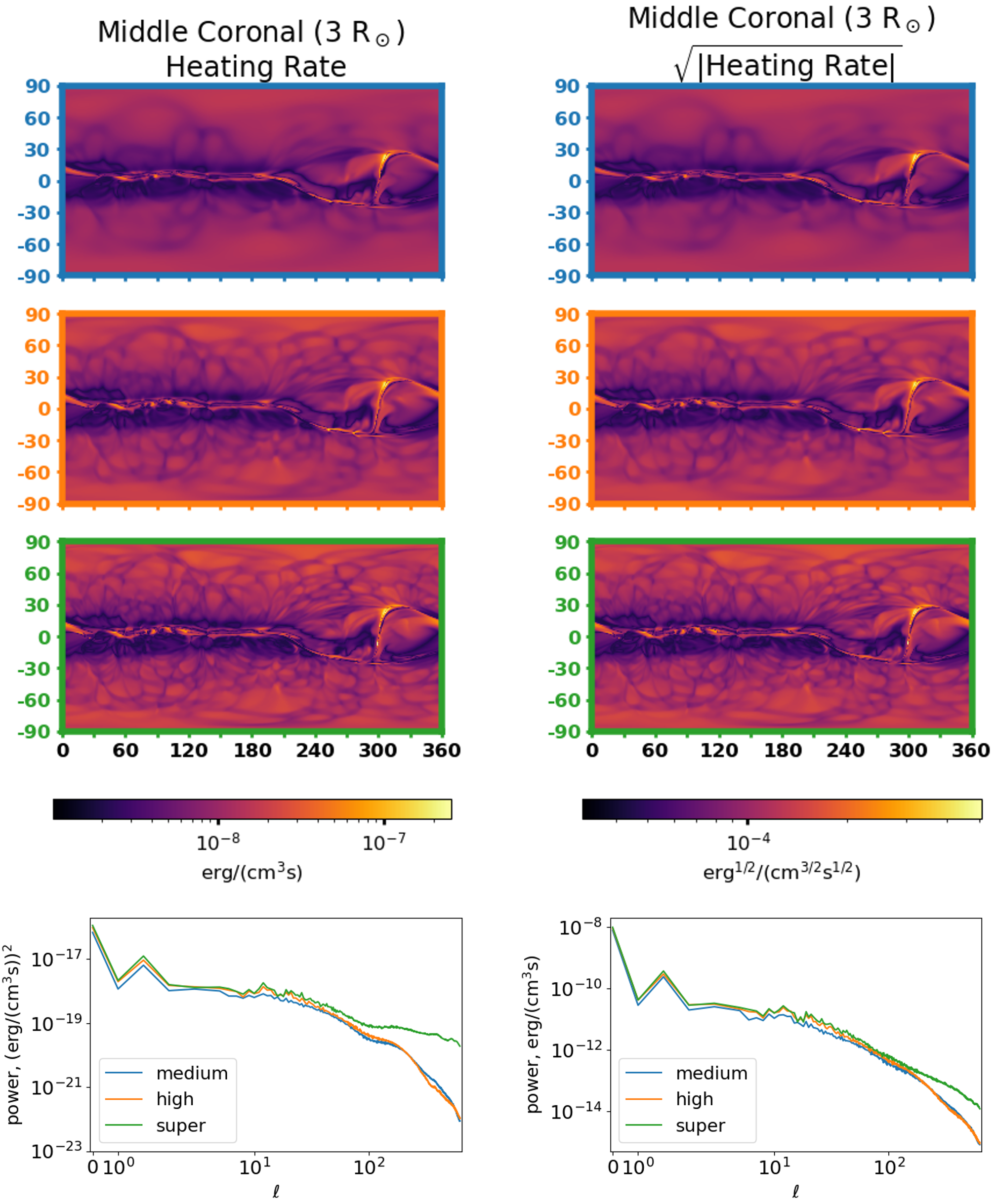}
    \caption{\resubi{Similarly to} Figure \ref{fig:heat_combo_1.03}, \resubi{we present} the visualization \resubi{and} total power spectra \resubi{of our coronal heating rate. Here, we select} a middle coronal (3 R$_\odot$) \resubi{height}. Again, the \resubi{greater magnitude of the super total power spectra (bottom row) demonstrates the increased} background heating \resubi{of the upper panels.} The structures extending from the HCS are more numerous in the super resolution\resubi{. The} separation of the power spectra\resubi{, including} at the highest $\ell$ values\resubi{,} quantifies \resubi{this}.}
   \label{fig:heat_combo_3}
\end{figure*}

\begin{figure*}[hbt!] 
   \centering
   \includegraphics[scale=0.41]{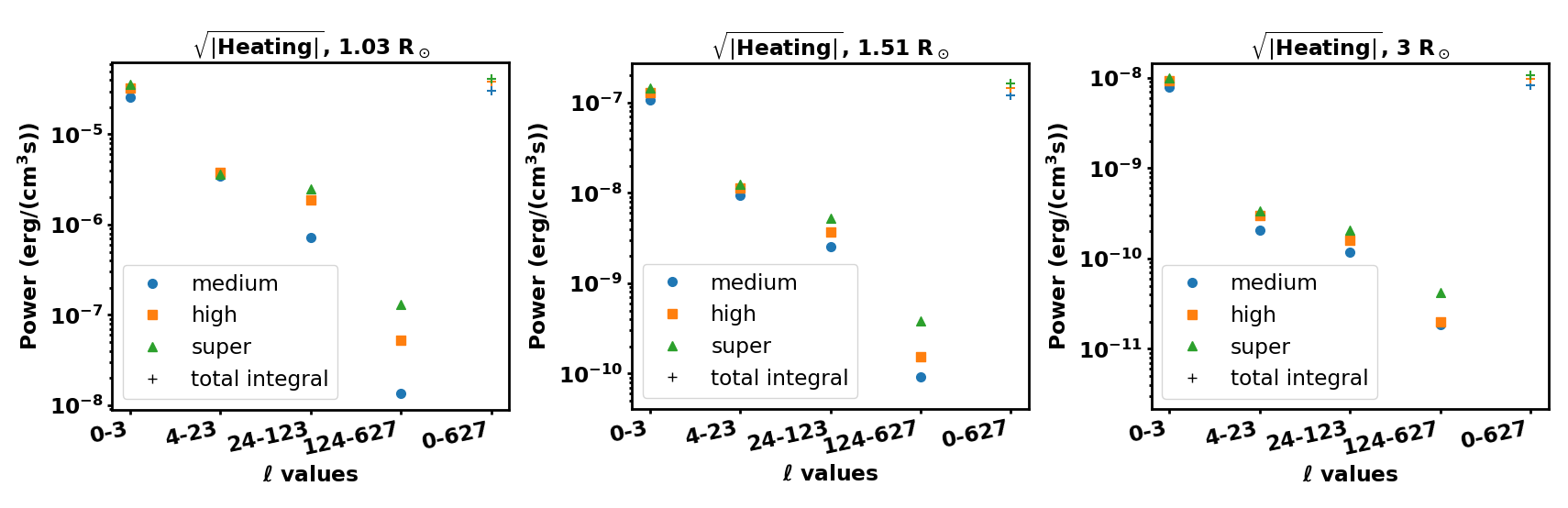}
    \caption{Similarly to Figures \ref{fig:br_slices} and \ref{fig:Q_slices}, this \resubi{presents power at given spatial scales separated into logarithmic-spaced bins. Here, we analyze the} square root of the absolute value of heating. As heating is not defined in the photosphere, we present only at the low coronal (1.03 R$_\odot$) and middle coronal (1.51 and 3 R$_\odot$) heights. We show 1.51 R$_\odot$ given its relevance to Table \ref{tab:vol_heating}. The amount of heating decreases in the middle corona as opposed to the low corona, but the differences in each bin \resubi{across} resolutions persists.}
   \label{fig:heat_slices}
\end{figure*}

We present the total power spectra of heating and the square root of its absolute value in the low corona (1.03 R$_\odot$) in Figure \ref{fig:heat_combo_1.03}. In the upper panels, we see increasing background values, particularly above $\pm 60 \degree$ latitude with increasing resolution. \resubi{With increasing resolution, there are more small-scale structures between these latitudes.}  This is reflected in the total power spectra of the bottom-most panels, where the power of the super simulation is generally higher than that of high or medium simulations across $\ell$ values.

In Figure \ref{fig:heat_combo_3}, we identically implement SHD for our middle corona\resubi{l (3 R$_\odot$)} slice. While the upper panels of Figure \ref{fig:heat_combo_1.03} had subtle background differences, the differences in background heating in the middle corona is more distinct. Further, in the super resolution, there are more structures branching away from the HCS along $\sim 0 \degree$ latitude as compared to the high and medium resolutions. 
This is again reflected in the total power spectra; across $\ell$ values, the magnitude of the super total power spectra exceeds that of high and medium.

Lastly, we summarize these results in Figure \ref{fig:heat_slices}. We use the same logarithmic bins as in Figures \ref{fig:br_slices} and \ref{fig:Q_slices}. We examine radial heights of 1.03, 1.51, and 3 R$_\odot$ to track heating from the low corona into the middle corona. In all three, the total power (denoted by the + marker in all three panels) of the super run is greater than both the total power of the high and medium runs. For 1.03 R$_\odot$, the second logarithmic bin ($\ell = [4,23]$) has similar values, but the high simulation has the most power. But by the onset of and into the middle corona (1.51 and 3 R$_\odot$), the super simulation once again has more power. For all heights, the higher $\ell$ value bins have distinct separation with the greatest power in the super simulation.

The increasing magnitude of the background heating with increasing resolution in both the low and middle corona suggests that increasing small-scale photospheric flux engenders heating across all spatial scales. This is seen in the super simulation having the most power for both the lowest $\ell$ value bin and overall heat throughout our radial slices. The comparatively much larger powers in the two highest $\ell$ value bins for the super simulation indicates that the small-scale flux generates more heating in the smallest spatial scales. However, the consistently higher power for the super simulation indicates that the simulation also contains more structures at every spatial scale. Therefore, the information of the photosphere magnetic flux (as shown in Figure \ref{fig:br_slices}) is likely entrenched in the simulation by the low corona.

\begin{deluxetable*}{cccccc}[hbt!]
 \tablecolumns{6}
\tablewidth{0pt}
\tablecaption{Total Volumetric Heating, 1.01 R$_\odot$ to 1.51 R$_\odot$ \label{tab:vol_heating}}
\tablehead{\colhead{} & \colhead{$\ell = [0, 3]$} & \colhead{$\ell = [4, 23]$} & \colhead{$\ell = [24, 123]$} & \colhead{$\ell = [124, 627]$} & \colhead{$\ell = [0, 627]$}}
 \startdata
 \quad & \quad & \quad & ergs/s & \quad & \quad \\
\hline
medium & 8.8459e+27 & 1.1917e+27 & 2.9181e+26 & 7.4692e+24 & 1.0337e+28\\
high & 1.0994e+28 & 1.3210e+27 & 6.8913e+26 & 3.3006e+25 & 1.3037e+28\\
super & 1.2208e+28 & 1.2738e+27 & 9.4306e+26 & 9.5021e+25 & 1.4520e+28 \\
\hline
 \quad & \quad & \quad & ratio & \quad & \quad \\
\hline
high/medium & 1.2428 &1.1086 & 2.3615  &4.4189 & 1.2612 \\
super/medium & 1.3800  & 1.06895  &3.2317 &12.7217  & 1.4046 
\enddata
\end{deluxetable*}

\begin{figure*}[hbt!] 
   \centering
   \includegraphics[scale=0.57]{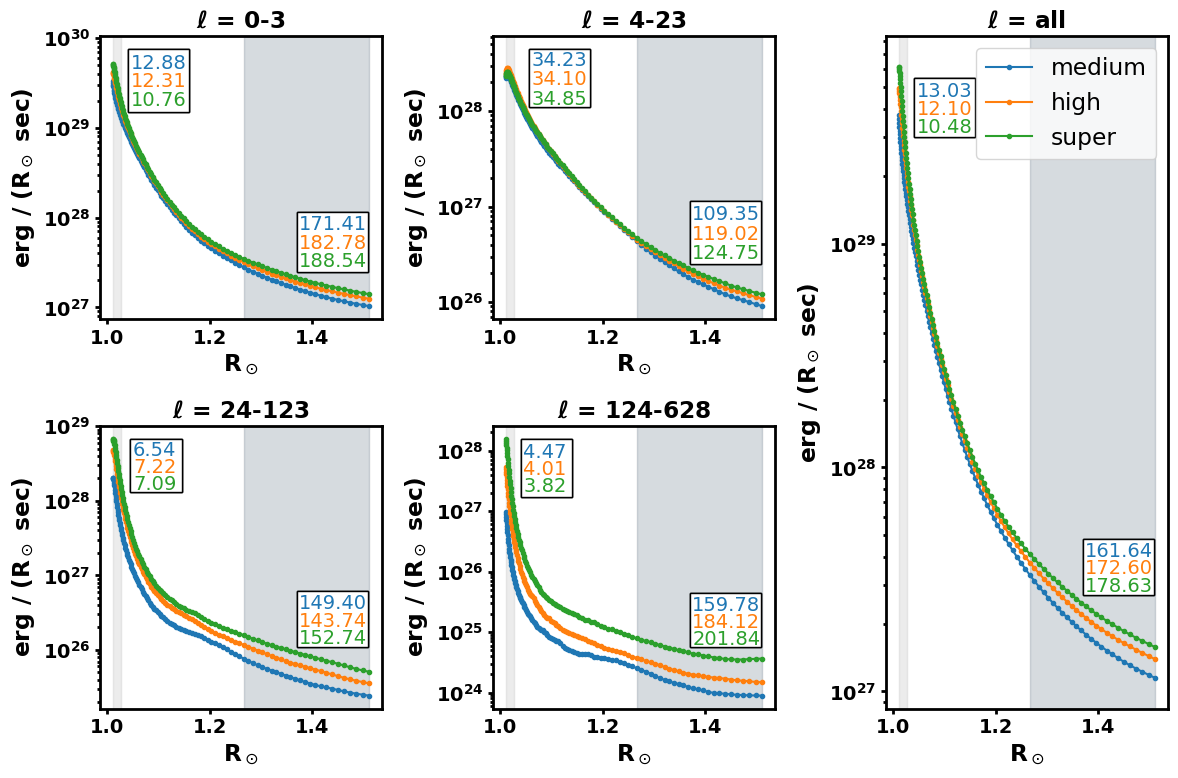}
    \caption{The four left panels indicate the amount of heat each radial slice contributes to the low corona, for each simulation (blue, medium; orange, high; green, super). Each represents one of our four logarithmic-spaced bins (top left: $\ell = [0, 3]$, top right: $\ell = [4, 23]$, bottom left: $\ell = [24, 123]$, and bottom right: $\ell = [124, 637]$). The larger right-most panel is the overall heat contribution to the volumetric heating rate ($\ell = [0, 627]$). All are plotted to show how each radial slice contributes to Table \ref{tab:vol_heating}. 
    \resubi{We perform exponential fits for every simulation of the form $y = Ae^{-r/\lambda}$ to characterize how heating evolves with height. For each panel, we give two $\lambda$ fits (in megameters) as the heating scale height in boxes. Each exponential fit is valid for either the data representing the onset of the low corona (lighter gray) or of the middle corona (darker gray). These fits characterize not only how heating evolves with height, but also how heating changes across our simulations.}} 
   \label{fig:heat_over_r}
\end{figure*}

Figures \ref{fig:br_slices} and \ref{fig:heat_slices} indicate that the  overall amount of heating deposited by the WTD model, i.e. the net Poynting flux injected at the inner boundary, evolves across resolutions. At 3 R$_\odot$, the power (in Gauss) in Figure \ref{fig:br_slices} is nearly identical in each logarithmic bin and in total. In Figure \ref{fig:heat_slices}, the power \resubi{at 3 R$_\odot$} (in erg/(cm$^{3}$ s)) of the super resolution is greater than the high and medium resolutions for every logarithmic bin and in total. This implies that more Poynting flux has propagated into the open flux tubes that form the solar wind in the super simulation. 
As quantified in \citet{Downs2016}, the balance between the outgoing Alfv\'en wave and its self-reflection determines the overall heating \citep[see also the work of][]{Sokolov2013}. This is mediated largely by the R$_2$ term in Equation \resubiv{(\ref{eqn:heat})}, which governs self-reflection and depends on the gradient of the Alfv\'en speed (as detailed in \S \ref{sec:mas}). 

Consequently, how the gradient of the Alfv\'en speed evolves with height plays a key role in determining the net Poynting flux into the flux-tube and thus the overall heating. In the two left-most panels in Figure \ref{fig:br_slices}, we see that the photospheric flux differences have all but disappeared in both the largest magnetic field structures and overall flux by the onset of the low corona (1.03 R$_\odot$). 

The only change \resubi{across} simulations is the resolution of the photospheric magnetic field\resubi{. This suggests that, on average,} the gradient of the Alfv\'en speed must differ \resubi{across} the three simulations between the photosphere and low corona. Both the precipitous drop in coronal densities from the base of the transition region and in the radial magnetic field before the low corona determine this gradient. With the additional photospheric flux in the super simulation, $|B|$ must decrease more dramatically than the high and medium cases, resulting in a shallower gradient in the Alfv\'en speed, which otherwise has a tendency to grow rapidly due to the sharp density gradient in the lower atmospheres. \resubii{A shallower Alfv\'en speed gradient (shallowest in the super simulation) leads to a smaller wave reflection term, enabling more of the input Poynting flux to propagate into the corona.} Therefore, the information of the photosphere critically determines the net Poynting flux and, accordingly, the net heat flux.

\subsection{Volumetric Heating in the Low Corona} \label{sec:vol}

Quantifying \resubi{how changing the }magnetic field \resubi{influences middle coronal heating} is critical, as the heat present in the middle corona sets the densities of the solar wind \citep{Parker1958}. We accordingly integrate radially over our discrete \resubii{angular} grid to yield \resubii{the} volumetric heating rate \resubi{of the low corona (in ergs/sec)}. We can also quantify how the different spatial scales contribute to the heating rate in each simulation (see Equation \resubiv{(\ref{parseval})}).

In Table \ref{tab:vol_heating}, we sum the total power spectra of $\sqrt{|\mathrm{heat}|}$ between 1.01 $R_\odot$ and 1.51 R$_\odot$. We use the same logarithmic\resubi{-spaced} bins to split the sum of the total power spectra into spatial regimes in the top three rows. In the two bottom-most rows, we include ratios to compare how much more heating exists in the super and high resolutions as compared to the medium resolution. While the magnitude of heating decreases for higher $\ell$ values, the high/medium and super/medium ratios increase with higher $\ell$ values. This indicates the presence of more small-scale structures as resolution increases.

We also present the total volumetric heating for all $\ell$ values in the rightmost column. There, we see that the super simulation has 40\% more heating than the medium simulation\resubi{.  Additionally,} the high simulation produces 26\% more heating. This close numeric match to the ratio of the first logarithmic bin ($\ell$ = [0,3]) indicates that the inclusion of more small-scale magnetic flux not only impacts every spatial scale, but particularly influences the heating of the largest scales. This is to say that the largest spatial scales of heating cannot be accurately captured without the presence of small-scale photospheric flux; it influences the propagation, reflection, and subsequent dissipation of wave energy along the flux tube.

This volumetric heating rate spans the low corona. Because the overall amount of heating depends on the resolution of the surface boundary condition, we likely can use this relationship to inject a heating correction into lower resolution simulations, where resolving the flux-distribution is not possible. This, in principle, would augment the results of computationally cheaper simulations to create a more physical parameter space during the critical transition to the middle corona. 

We also expect this result to be broadly useful for a more general class of wave-turbulence models or coronal heating models that depend on background coronal properties. WTD, as a subgrid model, is sensitive to the expansion of flux tubes and the magnetic field. \resubi{Alternative heating} models that nonetheless depend on these coronal loop properties should similarly see heating increases with the inclusion of small-scale photospheric flux.

\resubi{As opposed to the total values in Table \ref{tab:vol_heating}, we now examine how the average heating rate varies with height within the low corona. From Equation \resubiv{(\ref{parseval})}, we calculate the integral of the volumetric heating rate over area for each radial slice, spanning from 1.01 -- 1.51 $R_\odot$. In Figure \ref{fig:heat_over_r}, this integral is in the right-most panel. We also separate this integral into}  each of our \resubi{previously used} logarithmic-spaced \resubi{$\ell$ bins} in the four left panels. 

 \resubi{For all $\ell$ bins,} the amount of heat in the super simulation \resubi{generally} exceeds that of the high and the medium. \resubi{The $\ell = [4,23]$ bin presents a minor exception at the base of the low corona. This is not seen in any other bin or in the total heating in the rightmost panel.} As \resubi{the values of the }$\ell$ \resubi{bins} increase and therefore the spatial scale decreases, the separation in amount of heat across simulations increases.  

Both \resubi{the excess heating for the high simulation in the $\ell = [4,23]$ bin and increasing differences between simulations match the} values in Table \ref{tab:vol_heating}.\resubi{Unlike the integrated values in Table \ref{tab:vol_heating}, the profiles as a function of radius help us pinpoint where the differences in the integrated values begin to take hold. In particular, for $\ell>24$ we see that the offsets appear immediately at the base of the low corona. This further supports the idea that spatial information in the photospheric magnetic field structure becomes entrenched in the heating rate quite low down.}

Additionally, \resubi{for each panel of Figure \ref{fig:heat_over_r}, we} perform two separate exponential fits of the form $y = Ae^{-r/\lambda}$. \resubi{One fit is for} the \resubi{base} of the low corona (light gray\resubi{, spanning 1.01-1.02 R$_\odot$}) and the \resubi{other is for the top }of the \resubi{low} corona (dark gray\resubi{, for 1.27 - 1.51 R$_\odot$). The} value of $\lambda$\resubi{, as a heating scale height,} in megameters \resubi{is indicated} on each panel. 

In separating the heating by spatial scales, as opposed to by individual structures (e.g. see the work of \citet{Downs2016}), we can examine heating trends. The super heating initially decays more rapidly than the high and medium \resubi{heating. This is most distinct} in the smallest spatial scales, but also in the $\ell = [0,3]$ bin and overall. These steeper drop-offs match the expected result that high\resubiv{-}resolution structure at the base of the low corona is a key contributor to overall heating. However at the top of the low corona, across all spatial scales, the heating rate in the super simulation somewhat surprisingly decays more slowly than in the high and medium simulations. This reinforces the notion that the inclusion of small-scale photospheric magnetic flux has subtle influence on the heating properties \resubi{at the onsets of the low and middle corona.} 

\resubi{We note a hook at the base of the low corona for all simulations in the $\ell = [4,23]$ bin. As a result, the fitted} $\lambda$ values\resubi{ are more similar, as they represent an average across radial slices as opposed to capturing the trend of successive radial slices. 
While $\ell$-values do not indicate where the structures of a given size exist, the recomposition using different $\ell$ bins presented in Figure \ref{fig:br_SHViz_1} suggests that the active region strongly influences the the $\ell = [4,23]$ bin. This is consistent with heating within an active region differing from the quiet Sun, which could explain why the heating profile at the base of the low corona is distinct in this bin. Regardless, these fits are useful to characterize how the decay of heating differs at the base and top of the low corona. This information could be used to improve empirical heating prescriptions within models.}

\subsection{Relating Magnetic and Heating Structures}

\begin{figure*}[hbt!] 

\gridline{\fig{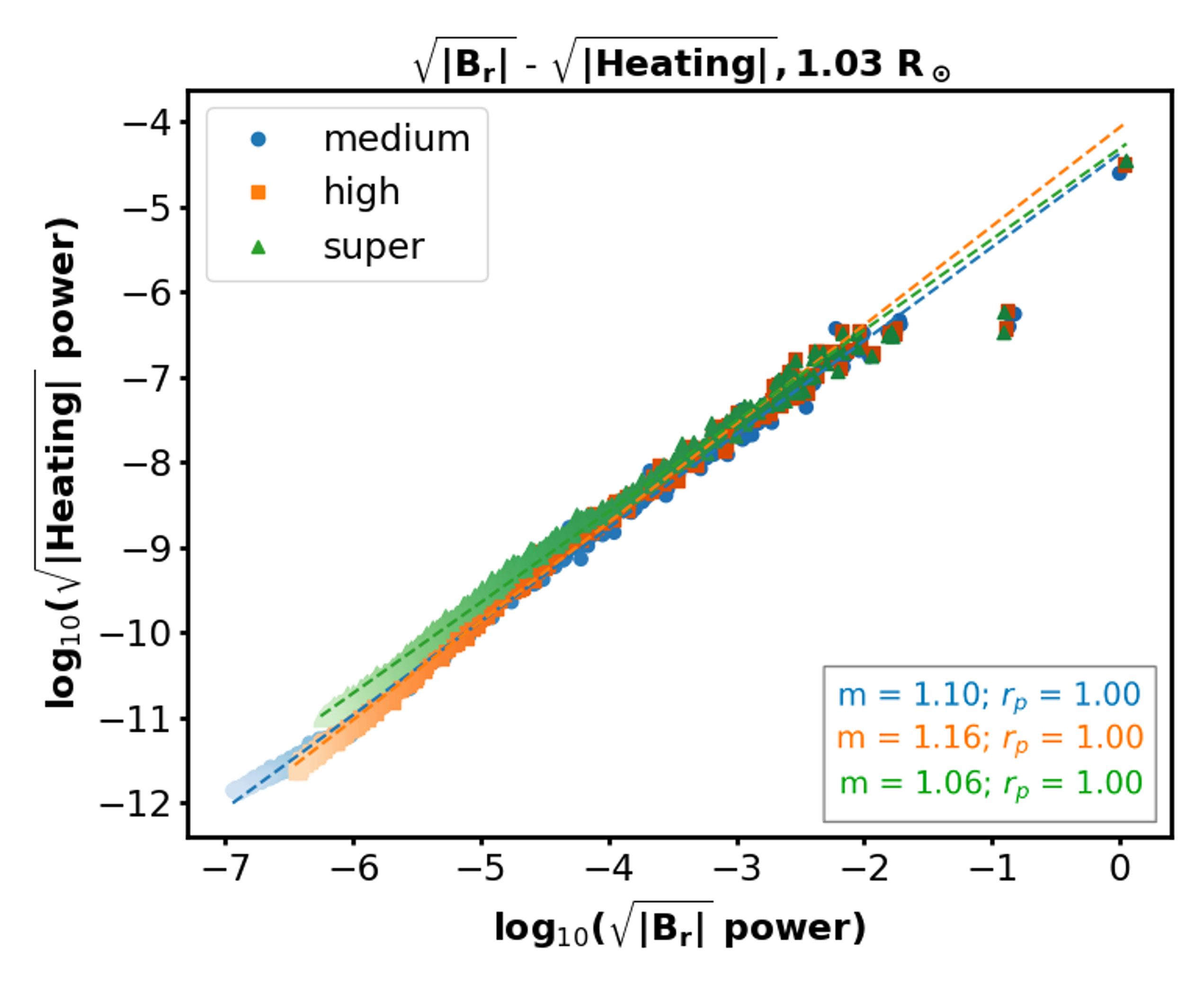}{0.33\textwidth}{(a)}
          \fig{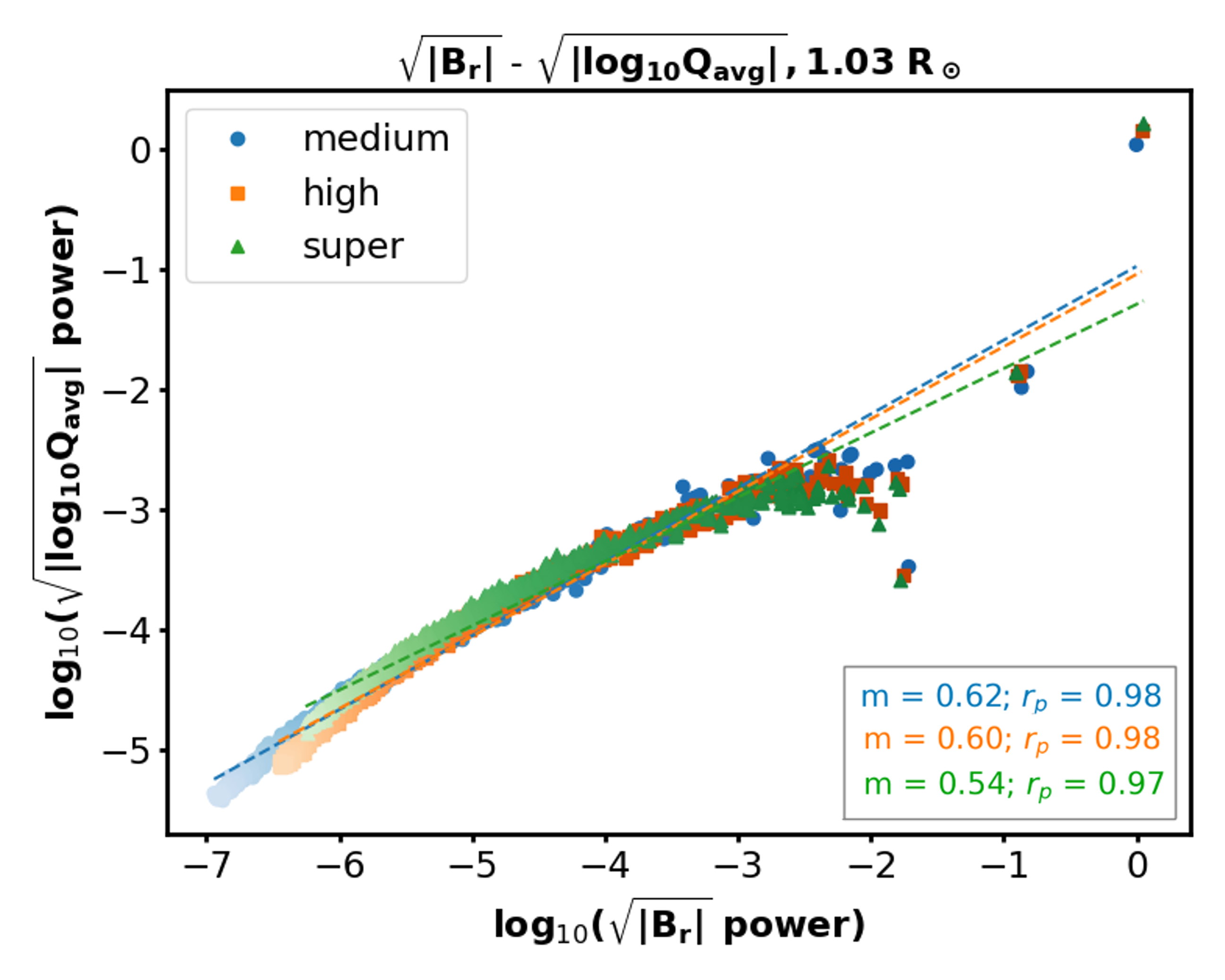}{0.33\textwidth}{(b)}
          \fig{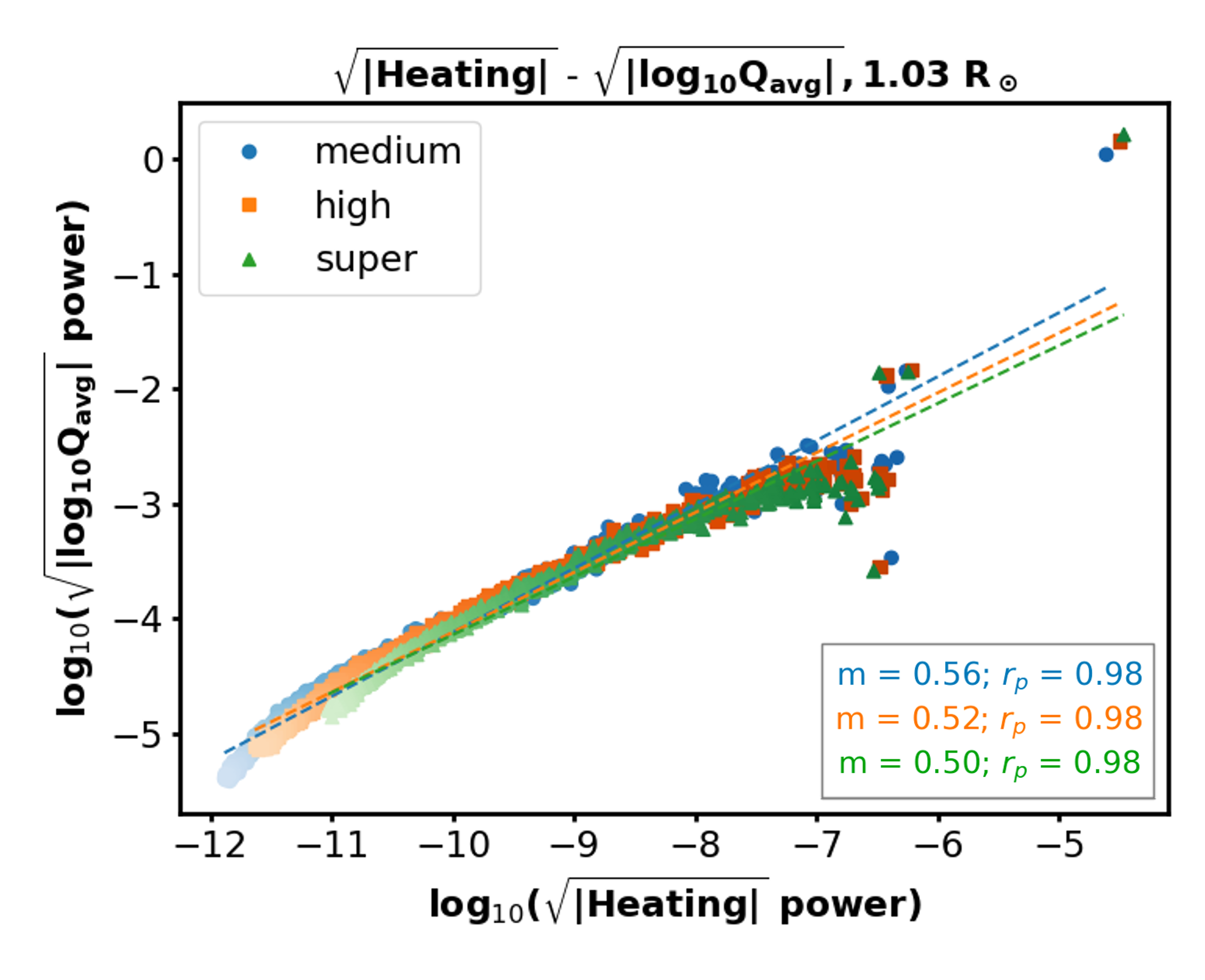}{0.33\textwidth}{(c)}}          
\caption{Comparisons of the power spectra of the square root of the absolute value of our three quantities of interest: the radial magnetic field, the heating, and the squashing factor. In the lower right-hand corner of each plot, we display the slope generated from linear regression and the \resubi{calculated} Pearson's r. In Panel (a), the 1:1 correlation of the radial magnetic field and heating holds true throughout the low corona. Panels (b) and (c) do not have a 1:1 relationship; this is due to the geometric and physical properties of the squashing factor. Nevertheless, heating and the squashing factor, as well as the radial magnetic field and squashing factor, are related quantities (as shown by their Pearson's r values). }
\label{fig:pvp}
\end{figure*} 

To confirm that photospheric magnetic flux is a key component in coronal heating, we compare the structures in magnetic flux, the squashing factor, and heating. Plotting the total power spectrum of one quantity at a given height versus another provides insight into the shapes of each. If the total power spectra are extremely similar, a plot in log-log space should have a slope around one. This indicates that the relative influence of spatial scales of each power spectrum matches, even if the power present in those spatial scales differ.

In Figure \ref{fig:pvp}, we present each combination of plotting the power spectra of the square root of the absolute value of the radial magnetic field, heating, and squashing factor against the power spectra of the other variables. To maintain information of the corresponding $\ell$ values, the darkest color values correspond to the lowest $\ell$ values and therefore the largest spatial structures (and still with blue, orange and green being the medium, high, and super resolutions, respectively). We also compute linear regression fits and Pearson's r ($\text{r}_{\text{P}}$) values.

In Panel (a), we analyze the shapes of the square root of the unsigned flux and square root of the heating for 1.03 R$_\odot$. At this radius and throughout the low corona, where the majority of the heating happens, the slope remains approximately 1; the slopes exhibit neither a consistent inverted order (medium greater than high greater than super) nor regular order (super greater than high greater than medium) hierarchy. This \resubi{indicates} that the relevant spatial scales in heating \resubi{match those of the} radial magnetic field\resubi{, regardless of resolution. This holds true for the} the low corona. 

\resubi{Panel (a) }quantifies the qualitative work that \citet{Downs2016} present for the 1D implementation of this heating schema. \resubi{For the radial magnetic field and heating, there is} an explicit correlation of the quantity of structures at every spatial scale, regardless of resolution. The slope being around 1 throughout the low corona indicates that the relative importance of structures in the \resubi{(square root of the)} unsigned magnetic flux and heating match. This power law relationship for the structure of the magnetic field and heating mimics the well-known coronal loop heating relationship to the magnetic field (see \citet{Mandrini2000} and references therein).

In the other panels of Figure \ref{fig:pvp}, we examine the relationships involving the squashing factor. The relationship between the magnetic field and its measure of complexity, in panel (b), is shallower than the slope of 1 in panel (a). The relationship between heat and the squashing factor, in panel (c), is shallower still. 

The shallowness of these slopes is an indicator of the nature of the squashing factor. The squashing factor is a topological measure that defines the boundary between discrete macroscopic magnetic flux domains in which the magnetic field connectivity is the same (i.e. similar plasma conditions). In contrast, the radial magnetic field and heating are more continuous quantities. Individual flux tubes, heated by braiding, turbulence, or both, cannot be resolved. In this sense, the departure in the shapes of the power spectra, and therefore the \resubi{differing import of particular spatial scales, is expected. O}ne-to-one correlation would only occur at the resolution limit of individual flux tubes. 


\section{Conclusions}\label{ref:conclusions}

We examined structures in the magnetic field, squashing factor Q, and heating rate for three global simulations of the corona during the July 2, 2019 total solar eclipse. Each 3D simulation (medium, high, and super) had identical parameterization except for the resolution of the HMI synoptic map, which changed via flux-preserving smoothing to our desired resolution. We then analyzed the impact of small-scale magnetic flux \resubi{on those quantities across spatial scales}.

Using spherical harmonic decomposition, we generated a total power spectrum for each region and quantity of interest. The total power spectrum (a function of $\ell$ only and not $m$) indicates the amount of structure at a given spatial scale (to which $\ell$ is inversely related). This set of power spectra allows for direct comparison across resolution, physical quantities of interest, and radial heights in the photosphere, low corona, and middle corona. Using Parseval's Theorem, we examined the quantity of structure in logarithmic-spaced spatial regimes. We calculated volumetric heating rates and quantified similarities of the structures of our data of interest. We offer the following conclusions.

(1) The inclusion of small-scale photospheric flux elements generates quantifiable structural complexity in the middle corona. While the radial magnetic field is nearly identical \resubi{for all of our simulations} at the HCS, the structures of the squashing factor and coronal heating rate vary in our three simulations. This is a consequence of the increasingly fragmented connectivity network that induces differing hydrodynamic states; the connectivity apportions the resulting coronal magnetic field between ever smaller individual flux concentrations. As a corollary, this analysis confirms that small-scale photospheric structures do not impact the open flux, as expected \citep{WangSheeley2002}. 

(2) The volumetric heating of the low corona is 40\% greater in the super simulation than the medium simulation. As the gradient of the Alfv\'en speed must rapidly change below the corona, we posit that the photosphere indelibly affects the heating of each \resubi{simulation}. In fact, we quantify that the heating is enhanced for every spatial scale of the corona. For the bin of the lowest $\ell$ values, the heating is 38\% greater, but for our highest $\ell$ values, the excess heating exceeds 1200\%. The near match in the additional power between the largest spatial scales (38\%) and the total heating (40\%) indicates the influence of small-scale features on overall heating.

(3) In the low corona, there is a 1:1 correlation between the structure of the unsigned magnetic flux and heating rate. This quantifies that our heating implementation depends on the magnetic field and that they share relative importance of spatial scales. \resubi{While not 1:1, there} nonetheless still exists a strong correlation between the squashing factor and both the unsigned magnetic flux and heating rate in the low corona. \resubi{As the squashing factor represents macroscopic division of magnetic flux domains, its spatial scales will be related but not perfectly matched as compared to the more continuous scales for the unsigned magnetic flux and heating.}

In tandem, these results underscore the influence of small-scale structure on global scales. \resubi{This study presents the opportunity to} generate \resubi{an empirical} correction \resubi{to capture unresolved features in low-resolution} global coronal models. \resubi{Such a correction could} improve accuracy without expending \resubi{undue computational resources. Further, it could} promote the unification of modeling efforts with cutting-edge, high-resolution observations that are now possible with e.g. the Daniel K. Inouye Solar Telescope \citep{Rimmele2020}. \resubi{An} ultra-high\resubiv{-}resolution magnetogram derived from an instrument such as Hinode/SOT \citep{Tsuneta2008} \resubi{could further refine this correction}.

\resubiii{It is worth noting that absolute uncertainty quantification of a computationally intensive 3D global coronal model remains an open research question in and of itself \citep[see e.g.][where only the last is for another MHD model]{Poduval2020, Issan2023, Jivani2023}. Standard methods like e.g. Monte Carlo methods for many free parameters would prove computationally prohibitive \resubiv{and an in-depth validation effort generally requires comparisons to many types of imaging and spectroscopic datasets. So, instead of providing direct comparisons to observations in this study, we leverage previous studies that have compared the high resolution model to white-light and emission line observations \citep[][]{Boe2021, Boe2022} and found reasonable agreement between 1 and 3 R$_\odot$. The total energy deposited is also on the same order of magnitude as previous coronal studies of global coronal heating for solar minimum conditions \citep[][though the exact numbers are strongly dependent on the full-sun flux-distribution at a given time]{Lionello2009, Downs2010}. Here}, our focus is on using controlled numerical experiments to isolate how photospheric resolution influences coronal structure. }

\resubiii{Nevertheless,} \resubi{to make any correction more broadly applicable, we would need additional resolution experiments. For example, our} results are based on a \resubi{fairly} quiescent Sun. \resubi{Moreover, we performed} our analysis \resubi{based on a single inner magnetic boundary condition}. Consequently, we could further develop our results with analysis of an active Sun and/or a time-dependent case \citep[e.g.][]{Downs2025}.

\resubi{We believe our results apply to any model sensitive to loop properties such as flux tube expansion and the magnetic field.} As a next step, we \resubi{seek to} implement \resubi{a} correction in future \resubi{low-resolution MAS runs}. This would enable broad reductions in \resubi{computational} time while increasing physical accuracy, expanding the use cases of MAS. Generalizing heating corrections to physics-informed models can provide valuable insight into the coronal heating problem and \resubi{solar wind acceleration}.

This work was supported by the University of Colorado Boulder George Ellery Hale Graduate Fellowship and the NASA Living With a Star program, in particular the Focused Science Topic: Towards a Quantitative Description of the Magnetic Origins of the Corona and Inner Heliosphere (grant 80NSSC22K1021). It was also supported by the NASA Heliophysics Supporting Research program (grant 80NSSC18K1129). Resources supporting this work were provided by the NASA High-End Computing (HEC) Program through the NASA Advanced Supercomputing (NAS) Division at Ames Research Center. We thank Maria Kazachenko, Bradley Hindman, Steven Cranmer, and Robert Ergun for their useful comments informing this work.

\bibliography{paperI}
\bibliographystyle{aasjournal}

\end{document}